\documentclass[11pt,letterpaper]{article}

\pdfoutput=1

\usepackage[utf8x]{inputenc}%
\usepackage{tcolorbox,fancybox}%
\usepackage{physics}
\usepackage{ntheorem}

\usepackage[english]{babel}
\usepackage{amsmath,amssymb,amsfonts}
\usepackage{graphicx}
\usepackage{booktabs}
\usepackage{color,xcolor}
\usepackage[numbers,sort&compress]{natbib}
\usepackage{tikz}
\usepackage{physics}
\usepackage{orcidlink}
\usepackage{slashed}
\usepackage{mathrsfs}
\usepackage{enumerate}
\usepackage{mdframed}
\usepackage{url}
\usepackage[makeroom]{cancel}
\usepackage{geometry}

\usepackage{pifont}

\newtheorem*{thm-non}{Theorem}

\usepackage{hyperref} 
\hypersetup{
    colorlinks=true,       
    linkcolor=red,          
    citecolor=blue,        
    filecolor=magenta,      
    urlcolor=purple           
}
\usepackage[all]{hypcap} 

\def\beqn{\begin{eqnarray}}
\def\eeqn{\end{eqnarray}}
\def\beqs{\begin{subequations}}
\def\eeqs{\end{subequations}}
\def\beq{\begin{equation}}
\def\eeq{\end{equation}}
\def\ba{\begin{array}}
\def\ea{\end{array}}

\def\non{\nonumber\\}

\def\hf{\frac{1}{2}}
\def\[{\left[}
\def\]{\right]}
\def\({\left(}
\def\){\right)}

\def\TeV{\rm TeV}
\def\GeV{\rm GeV}

\def\gSU{\rm SU}

\newcommand{\rep}[1]{\mathbf{#1}}
\newcommand{\repb}[1]{\mathbf{\overline{#1}}}

\def\Ac{\mathcal{A}}
\def\Bc{\mathcal{B}}

\def\Dc{\mathcal{D}}
\def\Ec{\mathcal{E}}

\def\Gc{\mathcal{G}}

\def\Lc{\mathcal{L}}
\def\Mc{\mathcal{M}}

\def\Oc{\mathcal{O}}

\def\Qc{\mathcal{Q}}

\def\Uc{\mathcal{U}}

\def\AG{\mathfrak{A}}  
\def\BG{\mathfrak{B}}  
\def\CG{\mathfrak{C}}

\def\XG{\mathfrak{X}}  
\def\YG{\mathfrak{Y}}

\title{
 {\bf Bottom quark and tau lepton masses in a toy ${\rm SU}(6)$ model } \\
\author{\large Ning Chen$^{\,\heartsuit}$\,\orcidlink{0000-0002-0032-9012}, Ying-nan Mao$\,^\star$\,\orcidlink{0000-0001-8063-8968}, \\ Zhaolong Teng$^{\,\clubsuit}$\,\orcidlink{0000-0002-7141-2331} }
\date{\small \it
$^\heartsuit\, ^\clubsuit $School of Physics, Nankai University, Tianjin, 300071, China \\
$^\star$ Department of Physics, School of Science, Wuhan University of Technology, \\ Wuhan, 430070, Hubei, China \\
}
}

\begin{document}

\maketitle
\setlength{\parskip}{0.2ex}

\begin{abstract}
\bigskip
We study a toy ${\rm SU}(6)$ model with the symmetry breaking pattern of the extended $331$ symmetry of ${\rm SU}(3)_c \otimes {\rm SU}(3)_W \otimes {\rm U}(1)_X$.
A ``fermion-Higgs mismatching'' symmetry breaking pattern is proposed for more realistic model building.
Within such symmetry breaking pattern, only one Higgs doublet develops vacuum expectation value for the spontaneous electroweak symmetry breaking, and gives tree-level top quark mass.
A natural VEV splittings in the $331$ breaking Higgs fields gives tree-level masses to both bottom quark and tau lepton.
The $125\,{\rm GeV}$ SM-like Higgs boson discovered at the LHC can have Yukawa couplings to bottom quark and tau lepton as in the SM prediction, and this suggests the $331$ symmetry breaking scale to be $\sim {\cal O}(10)\,{\rm TeV}$.
\end{abstract}

\vspace{9cm}
{\emph{Emails:}\\  
$^{\,\heartsuit}$\url{chenning_symmetry@nankai.edu.cn},\\
 $\,^\star$\url{ynmao@whut.edu.cn},\\
$^{\,\clubsuit}$ \url{tengcl@mail.nankai.edu.cn} }

\thispagestyle{empty}  
\newpage  
 
\setcounter{page}{1}  

\vspace{1.0cm}
\tableofcontents

\section{Introduction}
\label{sec:intro}
%
%

Grand Unified Theories (GUTs)~\cite{Georgi:1974sy,Fritzsch:1974nn} were proposed to unify all fundamental interactions and elementary particles described by the Standard Model (SM) at the electroweak (EW) scale.
Meanwhile, a unified description of the generational structure as well as the SM fermion mass hierarchies have not been realized in terms of the ${\rm SU}(5)$ or ${\rm SO}(10)$ GUTs.
This is largely due to the fact that three generations of SM fermions are accommodated in the ${\rm SU}(5)$ or ${\rm SO}(10)$ GUTs by simple repetition of one anomaly-free fermion generation.
Consequently, the symmetry breaking patterns do not provide any source for the observed SM fermion mass hierarchies.
It was pointed out and discussed in Refs.~\cite{Georgi:1979md,Frampton:1979cw,Frampton:1979tj,Barr:2008pn,Barr:2008gz} that multiple fermion generations, such as $n_g=3$ for the SM case, can be embedded non-trivially in GUT groups of ${\rm SU}(7)$ and beyond~\footnote{Such a scenario is different from the flavors with certain flavor symmetries, and is named as ``flavors without flavor symmetries'', see Refs.~\cite{Barr:2008pn,Barr:2008gz} for the recent discussions on the fermion mass generation from non-minimal GUTs.}.
Therefore, it is natural to conjecture that the SM fermion mass hierarchies may originate from the intermediate symmetry breaking scale of some non-minimal GUT with ${\rm SU}(N\geq 7)$~\cite{Frampton:1979cw,Frampton:1979tj}.
Historically, the embedding of the SM generations as well as fermion mass hierarchies were studied in the context of technicolor and extended technicolor models~\cite{Dimopoulos:1979es,Eichten:1979ah,Farhi:1980xs,Hill:2002ap,Appelquist:2003hn}, where the symmetry breakings are due to the fermion bi-linear condensates.
Given the discovery of a single $125\,{\rm GeV}$ SM-like Higgs boson at the Large Hadron Collider (LHC)~\cite{ATLAS:2012yve,CMS:2012qbp} until now, it is pragmatic to revisit the flavor issue in the framework of GUTs, where the spontaneous symmetry breakings are achieved by the Higgs mechanism.
\bigskip

Besides of addressing the flavor puzzle, it was also pointed out that the non-minimal GUTs can automatically give rise to the global Peccei-Quinn (PQ) symmetry~\cite{Peccei:1977hh} for the strong CP problem~\footnote{The first example was based on the rank-$2$ ${\rm SU}(9)$ theory~\cite{Georgi:1981pu}, where the fermion content of $ [5\times \repb{9_F}] \oplus \rep{36_F}$ enjoys the ${\rm SU}(5) \otimes {\rm U}(1)$ global symmetry.}.
This is due to the emergent global symmetry of ${\rm SU}(N)\otimes {\rm U}(1)$ in the rank-$2$ anti-symmetric ${\rm SU}(N+4)$ gauge theories (with $N\geq 2$), which was first pointed out by Dimopoulos, Raby, and Susskind~\cite{Dimopoulos:1980hn}.
In this regard, the longstanding flavor puzzle as well as the PQ quality problem~\cite{Barr:1992qq,Kamionkowski:1992mf,Holman:1992us} may be simultaneously addressed within the non-minimal GUTs~\footnote{Recently, it was realized~\cite{Chen:2021ovs} that the non-minimal GUTs with $n_g=3$ can generally lead to sufficiently large dimensional PQ-breaking operators for the later problem.}.
\bigskip

Before the ambitious goal of understanding the known SM fermion mass hierarchies in realistic non-minimal GUTs, it will be useful to ask whether the minimal version of this class already had some general properties in producing the SM fermion masses.
Among various non-minimal GUTs with ${\rm SU}(N\geq 7)$, indeed, an extended gauge symmetry of $\Gc_{331}\equiv {\rm SU}(3)_c \otimes {\rm SU}(3)_W \otimes {\rm U}(1)_X$ above the EW scale is usually predicted~\footnote{A specific example will be given for the ${\rm SU}(9)$ GUT in Sec.~\ref{subsec:SU8}.}.
This class of models are collectively known as the $331$ model, and were previously studied in Refs.~\cite{Lee:1977qs,Lee:1977tx,Pisano:1992bxx,Foot:1992rh,Montero:1992jk,Ng:1992st,Liu:1993gy,Pal:1994ba,Long:1995ctv,Tonasse:1996cx,Ponce:2002sg,Dias:2003zt,Dias:2004dc,Ferreira:2011hm,Dong:2012bf,Buras:2012dp,Machado:2013jca,Buras:2013dea,Boucenna:2014ela,Boucenna:2014dia,Boucenna:2015zwa,Deppisch:2016jzl,Cao:2016uur,Li:2019qxy,Chen:2021haa,CarcamoHernandez:2021tlv,Buras:2021rdg,Hernandez:2021zje,Alves:2022hcp}.
This motivates us to consider the ${\rm SU}(6)$ as a one-generational toy model~\footnote{The ${\rm SU}(6)$ turns out to be a toy model for the third generation in particular.}, which can be spontaneously broken to $\Gc_{331}$ by its adjoint Higgs field of $\rep{35_H}$.
An advantage of considering the one-generational ${\rm SU}(6)$ instead of the 331 model is that one can uniquely define the electric charges for both fermions and gauge bosons in the spectrum.
Meanwhile, the previous studies based on the 331 model itself often allowed different charge quantization schemes~\cite{Pisano:1992bxx,Buras:2012dp,Buras:2013dea,Cao:2016uur,Buras:2021rdg}, which could potentially lead to fermions with exotic electric charges.
\bigskip

After the GUT symmetry breaking, there can be three ${\rm SU}(3)_W$ anti-fundamental Higgs fields in the $331$ model.
In the previous studies, only one of them developed a vacuum expectation value (VEV) of $V_{331}$ for the symmetry breaking of $ {\rm SU}(3)_W \otimes {\rm U}(1)_X \to {\rm SU}(2)_W \otimes {\rm U}(1)_Y$, while two others developed VEVs of $v_{\rm EW} \simeq 246\,\GeV$ for the electroweak symmetry breaking (EWSB).
According to the Yukawa couplings, one can identify a type-II two-Higgs-doublet model (2HDM) at the EW scale for the $331$ model.
By extending to larger non-minimal GUTs for $n_g=3$, such as ${\rm SU}(9)$ as our example, the conventional symmetry breaking pattern in the $331$ model predicts more than two EW Higgs doublets.
This is certainly very problematic given that the direct searches for the second Higgs doublet at the Large Hadron Collider (LHC) give null results so far.
Motivated by the general features in the Higgs sector of the non-minimal GUT, we study the alternative symmetry breaking pattern with only one EW Higgs doublet coming from the $\rep{15_H}$ of the ${\rm SU}(6)$.
An immediate question is how do bottom quark and tau lepton acquire masses given the vanishing tree-level Yukawa couplings.
It turns out their masses can only be obtained when two ${\rm SU}(3)_W$ anti-fundamental Higgs fields from $\repb{6_H}^{\rho=1\,,2}$ develop VEVs both for the $331$ and EW symmetry breaking directions.
A natural mass splitting between the top quark and the $(b\,,\tau)$ in the third generation can be achieved with $\Oc(1)$ Yukawa couplings.
The corresponding $331$ breaking scale is found to be $V_{331} \sim \Oc(10)\,\TeV$ from the Yukawa couplings of the SM-like Higgs boson with the $(b\,,\tau)$.
Historically, a universal $\Oc(1)$ Yukawa coupling was motivated by observing the natural top quark mass at the EW scale, and this was generalized as the anarchical fermion mass scenario in the studies of the neutrino masses~\cite{Hall:1999sn,Haba:2000be}.
We also wish to remind the readers, the whole discussions are based on the $331$ model due to the minimal one-generational ${\rm SU}(6)$ symmetry breaking.
Aside from the SM fermion masses, we do not address some general questions of gauge coupling unification or proton lifetime predictions.
Neither do we determine whether a supersymmetric extension to the current model is necessary, with the belief that this would better be studied in more realistic models with $n_g=3$.
Some related discussions can be found in Refs.~\cite{Boucenna:2014dia,Deppisch:2016jzl,Chen:2021haa}.
\bigskip

The rest of the paper is organized as follows.
In Sec.~\ref{sec:SU6_mini}, we motivate the possible symmetry breaking pattern from two independent aspects in the toy ${\rm SU}(6)$ model, which leads to only one Higgs doublet for the spontaneous EWSB.
In Sec.~\ref{sec:SU6_Higgs}, we describe the Higgs sector in the ${\rm SU}(6)$ GUT, with the emphasis on the reasonable mass generations to the bottom quark and tau lepton through the Yukawa couplings.
In Sec.~\ref{sec:btau}, we describe the bottom quark and tau lepton masses in the toy ${\rm SU}(6)$ model based on the reasonable symmetry breaking pattern as well as the VEV assignment.
Some comments will be made for the necessary condition of the radiative mass generation in the current context.
We summarize our results and make discussions in Sec.~\ref{sec:conclusion}.
An appendix~\ref{sec:331_gauge} is provided to summarize the gauge sector as well as the fermion Yukawa couplings of the $331$ model.
All Lie group calculations in this work are carried out by {\tt LieART}~\cite{Feger:2012bs,Feger:2019tvk}.
\bigskip

\section{One-generational ${\rm SU}(6)$}
\label{sec:SU6_mini}

The minimal anomaly-free ${\rm SU}(6)$ GUT contains the left-handed fermions of 
\beqn
\{ f_L \}_{ {\rm SU}(6) } &=& \repb{6_F}^\rho \oplus \rep{15_F} \,,\quad \rho =1\,,2 \,.
\eeqn
The fermion sector enjoys a global symmetry of 
\beqn\label{eq:SU6_flavor}
{\cal G}_{\rm flavor}&=& {\rm SU}(2)_F \otimes {\rm U}(1)_{\rm PQ} \,,
\eeqn
according to Ref.~\cite{Dimopoulos:1980hn}.
The most general Yukawa couplings that are invariant under the gauge symmetry are expressed as~\footnote{Other Yukawa couplings of $\epsilon_{\rho \sigma } \repb{6_F}^\rho \repb{6_F}^\sigma ( \rep{15_H} +\rep{21_H})  + H.c.$ are also possible.
These terms are only relevant to neutrino masses, and we will neglect them in the current discussions.}
\beqn\label{eq:SU6_Yukawa}
-\Lc_Y&=& (Y_\Dc)_{\rho \sigma } \rep{15_F}  \repb{6_F}^\rho \repb{6_H}^\sigma  +  Y_\Uc \rep{15_F} \rep{15_F}  \rep{15_H}   + H.c. \,,
\eeqn
where we allow the explicit ${\rm SU}(2)_F$-breaking term in the Yukawa couplings, so that $(Y_\Dc)_{\rho \sigma} \neq Y_\Dc \delta_{\rho\sigma }$.
\bigskip

Below, we motivate our Higgs VEV assignments for the viable symmetry breaking from three different aspects, which include
\begin{itemize}

\item[$\AG$] the null results in searching for a second Higgs doublet at the LHC,

\item[$\BG$] the extension to the non-minimal GUTs with $n_g=3$, e.g., the ${\rm SU}(8)$ GUT~\cite{Barr:2008pn,Barr:2008gz},

\item[$\CG$]  the natural mass generation of the bottom quark and tau lepton with Yukawa couplings of $\sim \Oc(1)$.

\end{itemize}
\bigskip

\subsection{The symmetry breaking pattern}

The viable ${\rm SU}(6)$ breaking pattern is expected to be
\beqn\label{eq:SU6_pattern}
&& {\rm SU}(6) \xrightarrow{\Lambda_{\rm GUT} } {\cal G}_{331}  \xrightarrow{ V_{331}}  {\cal G}_{\rm SM}  \,,\non
&& {\cal G}_{331} = {\rm SU}(3)_c \otimes {\rm SU}(3)_W \otimes {\rm U}(1)_X \,,\quad  {\cal G}_{\rm SM}= {\rm SU}(3)_c \otimes {\rm SU}(2)_W \otimes {\rm U}(1)_Y \,,
\eeqn
where the GUT scale symmetry breaking is achieved by an ${\rm SU}(6)$ adjoint Higgs field of $\rep{35_H}$.
The ${\rm U}(1)_X$ charge for the $\rep{6} \in {\rm SU}(6)$ and ${\rm U}(1)_Y$ charge for the $\rep{3}_W \in {\rm SU}(3)_W$ are defined by
\beqs
\beqn
X( \rep{6} ) &\equiv& {\rm diag} ( -\frac{1}{3}\,, -\frac{1}{3} \,, -\frac{1}{3} \,, +\frac{1}{3} \,, +\frac{1}{3}\,, +\frac{1}{3}) \,,\label{eq:Xcharge_331}\\
 Y( \rep{3}_W )&\equiv& {\rm diag} ( \frac{1}{6} + X\,, \frac{1}{6} +X \,, -\frac{1}{3} + X) \,.\label{eq:Ycharge_331}
\eeqn
\eeqs
The electric charge operator of the ${\rm SU}(3)_W$ fundamental representation is expressed as a $3\times 3$ diagonal matrix
\beqn\label{eq:Qcharge_331}
Q ( \rep{3}_W )&\equiv& T_{{\rm SU}(3)}^3 + Y = {\rm diag} ( \frac{2}{3}+X \,,- \frac{1}{3}+X \,, -\frac{1}{3}+X )\,.
\eeqn 
with the first ${\rm SU}(3)$ Cartan generator of 
\beqn
T_{{\rm SU}(3)}^3&=& \frac{1}{2} {\rm diag} (1\,,-1 \,,0 ) \,.
\eeqn
\bigskip

Accordingly, we find that Higgs fields in Eq.~\eqref{eq:SU6_Yukawa} are decomposed as follows
\beqn\label{eq:Higgs_Br01}
&& \repb{6_H}^\rho = ( \repb{3}\,,\rep{1}\,, +\frac{1}{3} )_{ \mathbf{H}}^\rho \oplus \overbrace{ ( \rep{1}\,,\repb{3}\,, -\frac{1}{3} )_{ \mathbf{H} }^\rho}^{ \rep{\Phi}_{ \repb{3}\,, \rho} } \,,\non
&& \rep{15_H} = ( \repb{3}\,,\rep{1}\,, -\frac{2}{3} )_{ \mathbf{H}} \oplus \overbrace{ ( \rep{1}\,,\repb{3}\,, +\frac{2}{3} )_{ \mathbf{H}} }^{ \rep{\Phi}_{\repb{3}}^\prime } \oplus ( \rep{3}\,,\rep{3}\,, 0 )_{ \mathbf{H}} \,, 
\eeqn
for the symmetry breaking pattern in Eq.~\eqref{eq:SU6_pattern}.
Two $( \rep{ 1} \,, \repb{3} \,, -\frac{1}{3})_{ \mathbf{H}}^\rho \subset \repb{6_H}^\rho$ contain SM-singlet directions after the second stage symmetry breaking in Eq.~\eqref{eq:SU6_pattern}. 
Meanwhile, the $( \rep{ 1} \,, \repb{2} \,, +\frac{1}{2} )_{ \mathbf{H}} \subset ( \rep{ 1} \,, \repb{3} \,, +\frac{2}{3})_{ \mathbf{H}} \subset \rep{15_H}$ can only develop VEV to trigger the spontaneous EWSB of ${\rm SU}(2)_W \otimes {\rm U}(1)_Y \to {\rm U}(1)_{\rm em}$.
Under the symmetry breaking pattern in Eq.~\eqref{eq:SU6_pattern} and the charge quantization given in Eqs.~\eqref{eq:Xcharge_331}, \eqref{eq:Ycharge_331}, and \eqref{eq:Qcharge_331}, we summarize the ${\rm SU}(6)$ fermions and their names in Tab.~\ref{tab:SU6_331_ferm}.
For the SM fermions marked by solid underlines, we name them by the third generational SM fermions.
This will become manifest from their mass origin within the current context.
\bigskip

\begin{table}[htp]
\begin{center}
\begin{tabular}{c|c|c}
\hline \hline
   $\gSU(6)$   &  ${\cal G}_{331}$  & ${\cal G}_{\rm SM}$  \\
\hline \hline
  $\repb{6_F}^1 $ & $(\repb{ 3} \,, \mathbf{1} \,, + \frac{1}{3})_{ \mathbf{F} }^{1 } ~:~ (\Bc_R^{1 })^c$    &  $( \repb{ 3} \,, \mathbf{1} \,, + \frac{1}{3})_{ \mathbf{F} }^{1 }~:~ \underline{ {b_R}^c } $ \\   
    & $(\rep{1} \,, \repb{3} \,,  -\frac{1}{3})_{ \mathbf{F} }^{1} ~:~ \Lc_L^{1 }$  &  $( \rep{1} \,, \repb{ 2}  \,, -\frac{1}{2} )_{ \mathbf{F} }^{1 }~:~ \underline{ \ell_L = ( \tau_L \,, -\nu_L)^T} $\\ 
    &  & $( \mathbf{1} \,, \mathbf{ 1}  \,, 0 )_{ \mathbf{F} }^{1 }~:~ \widetilde N_{L}^1$  \\ \hline
 $\repb{6_F}^2$  & $(\repb{ 3} \,, \mathbf{1} \,, + \frac{1}{3})_{ \mathbf{F} }^{2} ~:~ (\Bc_R^{2})^c $   &  $( \repb{3}\,, \mathbf{1} \,, + \frac{1}{3})_{ \mathbf{F} }^{2 }~:~ { B_R}^c$ \\
    &  $(\mathbf{1} \,, \repb{ 3} \,,  -\frac{1}{3})_{ \mathbf{F} }^{2 } ~:~ \Lc_L^{2 }$ &  $( \rep{1} \,, \repb{ 2}  \,, -\frac{1}{2} )_{ \mathbf{F} }^{2 } ~:~ ( E_L \,, - N_L )^T $\\ 
   &   &  $( \mathbf{1} \,, \mathbf{1}  \,, 0 )_{ \mathbf{F} }^{2 }~:~  \widetilde N_L^2 $  \\ \hline
    $\rep{15_F}$ & $( \repb{3} \,, \mathbf{1} \,, - \frac{2}{3})_{ \mathbf{F}} ~:~ {t_R}^c$  & $( \repb{3} \,, \mathbf{1} \,, - \frac{2}{3})_{ \mathbf{F} }~:~ \underline{  {t_R}^c}$ \\ 
    & $( \mathbf{1}\,, \repb{3} \,, + \frac{2}{3})_{ \mathbf{F}} ~:~ ( \Ec_R)^c $  & $( \mathbf{1} \,, \repb{ 2}  \,, +\frac{1}{2})_{ \mathbf{F} }~:~ ( {N_R}^{c} \,, {E_R}^{c})^T $ \\ 
  &    & $( \mathbf{1} \,, \mathbf{1}  \,, +1)_{ \mathbf{F} }~:~\underline{ { \tau_R}^c} $ \\  
  & $(\mathbf{ 3} \,, \mathbf{3} \,, 0)_{ \mathbf{F} }~:~ \Qc_L$  &    $(\mathbf{ 3} \,, \mathbf{2} \,, +\frac{1}{6} )_{ \mathbf{F} }~:~ \underline{ q_L=( t_L\,, b_L)^T }$ \\  
  &   &   $(\mathbf{ 3} \,, \mathbf{1} \,, -\frac{1}{3} )_{ \mathbf{F} }~:~  B_L$ \\ \hline
\hline
\end{tabular}
\end{center}
\caption{
The $\gSU(6)$ fermion representations under the ${\cal G}_{331}$ and the ${\cal G}_{\rm SM}$.
All SM fermions are marked by solid underlines.}
\label{tab:SU6_331_ferm}
\end{table}%

In previous studies of the $331$ model~\cite{Lee:1977qs,Pisano:1992bxx,Foot:1992rh,Montero:1992jk,Ng:1992st,Liu:1993gy,Pal:1994ba,Long:1995ctv,Tonasse:1996cx,Ponce:2002sg,Dias:2003zt,Dias:2004dc,Ferreira:2011hm,Dong:2012bf,Buras:2012dp,Machado:2013jca,Li:2019qxy,Chen:2021haa,CarcamoHernandez:2021tlv,Buras:2021rdg,Hernandez:2021zje}, there were two Higgs doublets at the EW scale, which come from the $\repb{6_H}^1$ (chosen to be $\rho=1$ without loss of generality) and the $\rep{15_H}$, respectively.
Schematically, one expresses the VEVs of ${\rm SU}(3)$ anti-fundamentals as follows
\beqn\label{eq:331_VEVold}
&& \langle ( \rep{1}\,,\repb{3}\,, -\frac{1}{3} )_{ \mathbf{H}}^1   \rangle = \frac{1}{ \sqrt{2}} \left( \ba{c} 0 \\ v_d \\ 0 \\   \ea  \right) \,,\quad \langle ( \rep{1}\,,\repb{3}\,, -\frac{1}{3} )_{ \mathbf{H}}^2 \rangle= \frac{1}{ \sqrt{2}} \left( \ba{c} 0 \\ 0 \\ V_{331} \\   \ea  \right) \,, \non
&& \langle ( \rep{1}\,,\repb{3}\,, +\frac{2}{3} )_{ \mathbf{H}}  \rangle = \frac{1}{ \sqrt{2}} \left( \ba{c} v_u \\ 0 \\ 0 \\   \ea  \right) \,,
\eeqn
with $v_u^2 + v_d^2 = v_{\rm EW}^2 \approx (246\,{\rm GeV})^2$.
In such a scenario, the bottom quark and tau lepton masses are given by Yukawa couplings as follows
\beqn\label{eq:2HDMYuk_btau}
 \rep{15_F} \repb{6_H}^1 \repb{6_F}^1 +H.c. &\supset& \Big[  ( \rep{3}\,,\rep{3}\,, 0 )_{ \mathbf{F}} \otimes  ( \repb{3}\,,\rep{1}\,, +\frac{1}{3 } )_{ \mathbf{F}}^1  \non
&\oplus& ( \rep{1}\,,\repb{3}\,, +\frac{2}{3 } )_{ \mathbf{F}} \otimes ( \rep{1}\,,\repb{3}\,, -\frac{1}{3 } )_{ \mathbf{F}}^1 \Big]  \otimes  ( \rep{1}\,,\repb{3}\,, -\frac{1}{3 } )_{ \mathbf{H}}^1 + H.c. \non
&\supset&  \Big[ ( \rep{3}\,,\rep{2}\,, +\frac{1}{6} )_{ \mathbf{F}}   \otimes  ( \repb{3}\,,\rep{1}\,, +\frac{1}{3 } )_{ \mathbf{F}}^1 \non
&\oplus& ( \rep{1}\,,\rep{1}\,, +1)_{ \mathbf{F}}  \otimes ( \rep{1}\,,\repb{2}\,, -\frac{1}{2 } )_{ \mathbf{F}}^1 \Big]  \otimes  ( \rep{1}\,,\repb{2}\,, -\frac{1}{2} )_{ \mathbf{H}}^1 + H.c.  \,,
\eeqn
while the top quark mass is given by Yukawa coupling as follows
\beqn\label{eq:2HDMYuk_top}
\rep{15_F} \rep{15_F} \rep{15_H} + H.c. &\supset&  ( \rep{3}\,,\rep{3}\,, 0 )_{ \mathbf{F}} \otimes  ( \repb{3}\,,\rep{1}\,, -\frac{2}{3 } )_{ \mathbf{F}} \otimes  ( \rep{1}\,,\repb{3}\,, +\frac{2}{3 } )_{ \mathbf{H}} + H.c. \non 
&\supset& ( \rep{3}\,,\rep{2}\,, +\frac{1}{6} )_{ \mathbf{F}} \otimes ( \repb{3}\,,\rep{1}\,, -\frac{2}{3} )_{ \mathbf{F}}  \otimes ( \rep{1}\,,\repb{2}\,, +\frac{1}{2 } )_{ \mathbf{H}}   + H.c. \,.
\eeqn
All charged fermion masses in the third generation are due to the spontaneous EWSB.
Therefore, the Higgs sector at the EW scale is described by a type-II 2HDM~\footnote{This was also pointed out in Ref.~\cite{Chacko:2020tbu}, though a different symmetry breaking pattern was considered there.}.
Furthermore, the other Yukawa coupling of $ \rep{15_F} \repb{6_H}^2 \repb{6_F}^2 +H.c.$ give fermion masses of $m_B = m_E = m_N \sim {\cal O}(V_{331})$~\cite{Chen:2021haa}, according to the VEV assignment in Eq.~\eqref{eq:331_VEVold}.
By integrating out these massive fermions, the residual massless fermions are found to be anomaly-free under the ${\cal G}_{\rm SM}$.
Loosely speaking, one anti-fundamental fermion of $\repb{6_F}^2$ acquires mass with one Higgs field of $ ( \rep{1}\,,\repb{3}\,, -\frac{1}{3} )_{ \mathbf{H}}^2 \subset \repb{6_H}^2$ developing its VEV of $\Oc(V_{331})$~\footnote{In the current context, we do not discuss the mass origin for neutrinos of $\widetilde N_L^\rho \subset \repb{6_F}^\rho$. }.
Thus, we name such symmetry breaking pattern as the ``fermion-Higgs matching pattern''.
\bigskip

However, the ongoing probes of the second Higgs doublet at the LHC lack direct evidences for the predicted neutral and charged Higgs bosons from various channels~\cite{CMS:2019bfg,ATLAS:2019tpq,CMS:2019pzc,CMS:2019rlz,CMS:2019kca,CMS:2019ogx,CMS:2019bnu,CMS:2020imj,ATLAS:2020zms,ATLAS:2020tlo,ATLAS:2020gxx,ATLAS:2021upq}.
In the type-II 2HDM, hierarchical Yukawa couplings of $Y_\Uc \gg Y_\Dc$ can be expected for the third-generational SM fermion masses.
As will be shown below, the suppressed $(b\,,\tau)$ masses can be realized with a more natural Yukawa couplings of $Y_\Uc \sim Y_\Dc \sim \Oc(1)$ in the current context.
Aside from the experimental facts, it is most natural to consider the following VEVs for the Higgs fields
%
%
\beqn\label{eq:331_VEVnaive}
&& \langle ( \rep{1}\,,\repb{3}\,, -\frac{1}{3} )_{ \mathbf{H}}^{\rho=1\,,2}   \rangle = \frac{1}{ \sqrt{2}} \left( \ba{c} 0 \\ 0 \\ V_\rho \\   \ea  \right) \,,\quad \langle ( \rep{1}\,,\repb{3}\,, +\frac{2}{3} )_{ \mathbf{H}}  \rangle = \frac{1}{ \sqrt{2}} \left( \ba{c} v_u \\ 0 \\ 0 \\   \ea  \right) \,,
\eeqn
purely from the group theoretical point of view.
Obviously, two $\Gc_{331}$-breaking VEVs in Eq.~\eqref{eq:331_VEVnaive} are in the SM-singlet components, and one expects the natural hierarchy of $V_1 \sim V_2 \sim \Oc(V_{331} ) \gg v_u = v_{\rm EW}$.
By taking the ${\rm SU}(2)_F$-invariant Yukawa couplings of $ (Y_\Dc)_{\rho\sigma } = Y_\Dc  \delta_{\rho \sigma}$ in Eq.~\eqref{eq:SU6_Yukawa} for simplicity, we have the following mass terms
\beqn\label{eq:SU6Yukawa_SU2}
&& Y_\Dc \rep{15_F} \repb{6_H}^\rho \repb{6_F}^\rho + H.c. \supset  Y_\Dc  \Big[ ( \mathbf{3}\,, \mathbf{ 3}\,, 0 )_{ \mathbf{F}} \otimes  ( \mathbf{1}\,, \mathbf{ \bar 3}\,,  -\frac{1}{3} )_{ \mathbf{H}}^\rho  \otimes  ( \mathbf{\bar 3}\,, \mathbf{ 1}\,, +\frac{1}{3} )_{ \mathbf{F}}^{ \rho}   \non
&\oplus& (  \mathbf{ 1}\,, \mathbf{\bar 3}\,, +\frac{2}{3} )_{ \mathbf{F}} \otimes  ( \mathbf{1}\,, \mathbf{ \bar 3}\,,  -\frac{1}{3} )_{ \mathbf{H}}^\rho   \otimes (  \mathbf{ 1}\,, \mathbf{\bar 3}\,, -\frac{1}{3} )_{ \mathbf{F}}^{\rho} \Big] + H.c.   \non
&\Rightarrow & \frac{1}{ \sqrt{2}} (\bar b_L \,, \bar B_L ) \cdot \left( \ba{cc}  
0 &  0 \\
Y_\Dc V_1 &  Y_\Dc V_2 \\   \ea  \right) \cdot \left( \ba{c} b_R \\ B_R \\   \ea  \right) +\frac{1}{ \sqrt{2}}  (\bar \tau_L \,, \bar E_L ) \cdot \left( \ba{cc}  
0 &  -Y_\Dc V_1 \\
0 &  -Y_\Dc V_2 \\   \ea  \right) \cdot \left( \ba{c} \tau_R \\ E_R \\   \ea  \right) \non
& +& H.c. \,,
\eeqn
with the alternative VEV assignment in Eq.~\eqref{eq:331_VEVnaive}.
Clearly, both bottom quark and tau lepton remain massless after the spontaneous breaking of the $331$ symmetry.
Meanwhile, there is still only one anti-fundamental fermion of $\repb{6_F}^2$ becoming massive.
In this regard, the alternative symmetry breaking pattern achieved by both $ ( \rep{ 1} \,, \repb{3} \,, -\frac{1}{3})_{ \mathbf{H}}^\rho \subset \repb{6_H}^\rho$ is also valid from the anomaly-free condition.
Thus, we name the VEV assignment in Eq.~\eqref{eq:331_VEVnaive} as the ``fermion-Higgs mismatching pattern'' of symmetry breaking.
As we shall show below, this VEV assignment leads to a distinct Higgs spectrum from the 2HDM at the EW scale.
\bigskip

\subsection{An example: ${\rm SU}(8)$ with three generations}
\label{subsec:SU8}

Besides of the above phenomenological consideration, a better motivation of the current study can be made for non-minimal GUTs with multiple generations. 
Let us take the ${\rm SU}(8)$ GUT as an example, which can automatically lead to $n_g=3$ with the following fermion content~\cite{Barr:2008pn,Barr:2008gz}
\beqn\label{eq:SU8_ferm}
\{ f_L  \}_{ {\rm SU}(8) }&=& \[ \repb{8_F}^\rho   \oplus \rep{28_F} \] \bigoplus \[  \repb{8_F}^{\dot \rho}  \oplus \rep{56_F} \]  \,,
\eeqn
according to the rule in Ref.~\cite{Georgi:1979md}.
This setup enjoys an emergent global symmetry of $\[ {\rm SU}(4)_1 \otimes {\rm U}(1)_{1} \] \otimes \[ {\rm SU}(5)_2 \otimes {\rm U}(1)_{2} \]$~\cite{Dimopoulos:1979es,Chen:2021ovs}, with the flavor indices of $\rho=1\,,...\,,4$ and $\dot \rho=5\,,...\,,9$.
To focus on the third-generational fermions, we only consider the rank-$2$ sector in Eq.~\eqref{eq:SU8_ferm}.
${\rm SU}(4)_1$-invariant Yukawa couplings of $ \rep{28_F} \repb{8_H}^\rho \repb{8_F}^\rho + H.c.$ can be expected to give bottom quark and tau lepton masses, which are analogous to $ \rep{15_F} \repb{6_H}^\rho  \repb{6_F}^\rho + H.c.$ in the one-generational ${\rm SU}(6)$ model.
Another gauge-invariant Yukawa coupling of $\rep{28_F} \rep{28_F} \rep{70_H} + H.c.$ is expected to give the top quark mass.
A possible symmetry breaking pattern of ${\rm SU}(8)$ can be expected as follows
\beqn\label{eq:SU8_Pattern}
&& {\rm SU}(8) \xrightarrow{\Lambda_{\rm GUT} } {\rm SU}(4)_c \otimes {\rm SU}(4)_W \otimes  {\rm U}(1)_{X_0 } \xrightarrow{ V_{441} } {\rm SU}(3)_c \otimes {\rm SU}(4)_W \otimes  {\rm U}(1)_{X_1 }  \non
&\xrightarrow{  V_{341} }& {\rm SU}(3)_c \otimes {\rm SU}(3)_W \otimes  {\rm U}(1)_{X_2}  \xrightarrow{ V_{331} }  {\rm SU}(3)_c \otimes {\rm SU}(2)_L \otimes  {\rm U}(1)_Y \,.  
\eeqn
Here, $(V_{441} \,, V_{341} \,, V_{331})$ represent three intermediate symmetry-breaking scales above the EW scale.
The corresponding ${\rm U}(1)$ charges in Eq.~\eqref{eq:SU8_Pattern} are defined by
\beqs\label{eqs:SU8_charges}
\beqn
X_0(\rep{8}) &\equiv& {\rm diag} (  - \frac{1}{4} \,, - \frac{1}{4} \,, - \frac{1}{4}\,, - \frac{1}{4}  \,, +\frac{1}{4} \,, +\frac{1}{4} \,, +\frac{1}{4} \,, +\frac{1}{4}  ) \,,\label{eq:X0charge} \\
X_1 ( \rep{4}_c ) &\equiv& {\rm diag} ( - \frac{1}{12} + X_0 \,, - \frac{1}{12} + X_0 \,, - \frac{1}{12} + X_0\,,   \frac{1}{4} + X_0 )\,,\label{eq:X1charge}\\
X_2 ( \rep{4}_W ) &\equiv& {\rm diag} ( \frac{1}{12} + X_1 \,, \frac{1}{12} + X_1 \,, \frac{1}{12} + X_1 \,,  -\frac{1}{4} + X_1 )\,,\label{eq:X2charge}\\
Y ( \rep{4}_W )&\equiv&  {\rm diag} ( \frac{1}{4} + X_1 \,, \frac{1}{4} + X_1 \,, -\frac{1}{4} + X_1 \,, -\frac{1}{4} + X_1 )\,. \label{eq:Ycharge}
\eeqn
\eeqs
Following the above symmetry breaking pattern and charge quantizations in Eqs.~\eqref{eqs:SU8_charges}, one can decompose the minimal set of Higgs fields as
\beqs\label{eqs:SU8_HiggsDecomp}
\beqn
\repb{8_H}^{ \rho } &\supset& ( \repb{4} \,, \rep{1}\,, +\frac{1}{4})_{ \mathbf{H} }^\rho \oplus ( \rep{1} \,, \repb{4}\,, -\frac{1}{4} )_{ \mathbf{H} }^\rho  \,,\\
( \repb{4} \,, \rep{1}\,, +\frac{1}{4} )_{ \mathbf{H} }^\rho& =& ( \repb{3} \,, \rep{1}\,, +\frac{1}{3} )_{ \mathbf{H} }^\rho \oplus ( \rep{1} \,, \rep{1}\,, 0 )_{ \mathbf{H} }^\rho  \,,\\
( \rep{1} \,, \repb{4}\,, -\frac{1}{4} )_{ \mathbf{H} }^\rho& =& ( \rep{1} \,, \repb{3}\,, -\frac{1}{3})_{ \mathbf{H}}^\rho \oplus ( \rep{1} \,, \rep{1}\,, 0)_{ \mathbf{H} }^{\rho^\prime}  = ( \rep{1} \,, \repb{2}\,, -\frac{1}{2})_{ \mathbf{H}}^\rho \oplus ( \rep{1} \,, \rep{1}\,, 0)_{ \mathbf{H} }^\rho \oplus ( \rep{1} \,, \rep{1}\,, 0)_{ \mathbf{H} }^{\rho^\prime}  \,,\\
\rep{70_H}&\supset&  ( \rep{4} \,, \repb{4}\,, +\frac{1}{2})_{ \mathbf{H}}  \supset ( \rep{1} \,, \repb{4}\,, +\frac{3}{4})_{ \mathbf{H}} \non
&\supset& ( \rep{1} \,, \repb{3}\,, +\frac{2}{3})_{ \mathbf{H}}^{\prime \prime \prime} = ( \rep{1} \,, \repb{2}\,, +\frac{1}{2} )_{ \mathbf{H}}^{\prime \prime \prime} \oplus ( \rep{1} \,, \rep{1}\,, +1 )_{ \mathbf{H}}   \,.
\eeqn
\eeqs
Both the $\repb{8_H}^{ \rho }$ and the $\rep{70_H}$ contain the EWSB components of $( \rep{1} \,, \repb{2} \,, -\frac{1}{2})_{\mathbf{H}}^\rho$ and $( \rep{1} \,, \repb{2} \,, +\frac{1}{2})_{\mathbf{H} }$, respectively.
Besides, the four $\repb{8_H}^{ \rho }$ Higgs fields contain three singlet components for the intermediate symmetry breaking in Eq.~\eqref{eq:SU8_Pattern}.
A more careful counting by the anomaly-free condition at each stage of symmetry breaking shows that the Higgs spectrum is left with one EW Higgs doublets from the $\repb{8_H}^{ \rho }$ and one from the $\rep{70_H}$ if one adopted the ``fermion-Higgs matching pattern'' of symmetry breaking, as can be expected for the $n_g=3$ case.
In this regard, to have a realistic Higgs spectrum at the EW scale, a ``fermion-Higgs mismatching pattern'' of the intermediate symmetry breaking can be generally expected.
\bigskip

\section{The Higgs sector of the ${\rm SU}(6)$}
\label{sec:SU6_Higgs}

In this section, we describe the Higgs sector according to the symmetry breaking pattern in Eq.~\eqref{eq:SU6_pattern}, which consists of ${\rm SU}(3)_W$ anti-fundamentals of $\Phi_{ \repb{3}\,,\rho } \equiv ( \rep{1}\,, \repb{3} \,, - \frac{1}{3})_{ \mathbf{H}}^\rho \subset \repb{6}_{ \mathbf{H} }^\rho$ (with $\rho=1\,,2$) and $\Phi_{ \repb{3} }^\prime \equiv ( \rep{1}\,, \repb{3} \,, +\frac{2}{3})_{ \mathbf{H}} \subset \rep{15_H}$ after the ${\rm SU}(6)$ GUT symmetry breaking.
\bigskip

\subsection{The Higgs potential}

The most general ${\rm SU}(6)$ Higgs potential contains Higgs fields of $(\repb{6_H}^\rho \,, \rep{15_H}\,, \rep{35_H} )$.
The adjoint Higgs field of $\rep{35_H}$ is responsible for the GUT symmetry breaking of ${\rm SU}(6)\to \Gc_{331}$.
For our purpose, only the Higgs fields of $(\repb{6_H}^\rho \,, \rep{15_H})$ will be relevant for the sequential symmetry breakings.
At the GUT scale, the following terms can be expected in the Higgs potential
\beqn\label{eq:SU6_potential}
V&\supset& m_{11}^2 | \repb{6_H}^1 |^2 + m_{22}^2 |\repb{6_H}^2 |^2 - \Big( m_{12}^2 \repb{6_H}^1 ( \repb{6_H}^2 )^\dag + H.c.  \Big) + \frac{ \lambda_1}{2} | \repb{6_H}^1  |^4 + \frac{ \lambda_2}{2} | \repb{6_H}^2  |^4 \non
&+& \lambda_3 | \repb{6_H}^1 |^2 | \repb{6_H}^2  |^2 + \lambda_4 ( \repb{6_H}^{1\, \dag} \repb{6_H}^2 ) ( \repb{6_H}^{2\, \dag} \repb{6_H}^1 ) + \frac{\lambda_5 }{2 } \Big[ ( \repb{6_H}^{1\, \dag} \repb{6_H}^2 )^2 + H.c. \Big] \non
&+& m^2 |  \rep{15_H}  |^2 + \lambda |  \rep{15_H}  |^4 \non
&+& \Big( \kappa_1 | \repb{6_H}^1 |^2  + \kappa_2 | \repb{6_H}^2 |^2  \Big) |  \rep{15_H}  |^2 + \kappa_3 (\repb{6_H}^{1\, \dag} \rep{15_H} ) (\rep{15_H}^\dag \repb{6_H}^{1}  )  + \kappa_4 (\repb{6_H}^{2\, \dag} \rep{15_H} ) (\rep{15_H}^\dag \repb{6_H}^{2}  )   \non
&+& \Big(  \nu\,  \repb{6_H}^{[ 1} \repb{6_H}^{2] } \rep{15_H}  + H.c. \Big) \,.
\eeqn
The Higgs potential contains the mass squared parameters of $( m_{11}^2 \,, m_{22}^2 \,, m_{12}^2\,, m^2 )$, dimension-one parameter of $\nu$, and dimensionless self couplings of $(\lambda_{1\,,...\,,5}\,,\lambda \,, \kappa_{1\,,...\,,4})$.
After the GUT symmetry breaking, we assume all ${\rm SU}(3)_c$ colored components of $(\repb{6_H}^\rho \,, \rep{15_H})$ obtain heavy masses of $\Lambda_{\rm GUT}$.
The residual massless Higgs fields transforming under the ${\rm SU}(3)_W \otimes {\rm U}(1)_X$ symmetry form the following Higgs potential
\beqs\label{eqs:2HTM_potential}
\beqn
V_{\rm tot}&=& V( \Phi_{ \repb{3}\,, \rho } ) + V( \Phi_{ \repb{3}}^\prime ) + V( \Phi_{ \repb{3}\,, \rho}\,,\Phi_{ \repb{3}}^\prime )   \,,\\[1mm]
V( \Phi_{ \repb{3}\,, \rho } )&=& m_{11}^2 | \Phi_{ \repb{3}\,, 1 } |^2 + m_{22}^2 | \Phi_{ \repb{3}\,, 2 } |^2 - \Big( m_{12}^2 \Phi_{ \repb{3}\,, 1 }^\dag \Phi_{ \repb{3}\,, 2 } + H.c.  \Big) \non
&+& \frac{ \lambda_1}{2} | \Phi_{ \repb{3}\,, 1 }  |^4 + \frac{ \lambda_2}{2} | \Phi_{ \repb{3}\,, 2 }  |^4 + \lambda_3 | \Phi_{ \repb{3}\,, 1}|^2 |\Phi_{ \repb{3}\,, 2 }|^2 \non
&+& \lambda_4 (\Phi_{ \repb{3}\,, 1 }^\dag \Phi_{ \repb{3}\,, 2 } ) (\Phi_{ \repb{3}\,, 2 }^\dag \Phi_{ \repb{3}\,, 1 } ) + \frac{\lambda_5}{2} \Big[  (\Phi_{ \repb{3}\,, 1 }^\dag \Phi_{ \repb{3}\,, 2 } )^2  + H.c. \Big] \,,\label{eq:2HTM_potential01} \\[1mm]
V( \Phi_{ \repb{3}}^\prime )&=& m^2 | \Phi_{ \repb{3}}^\prime |^2 + \lambda | \Phi_{ \repb{3}}^\prime |^4 \,,\label{eq:2HTM_potential02} \\[1mm]
V( \Phi_{ \repb{3}\,, \rho}\,,\Phi_{ \repb{3}}^\prime ) &=& \Big( \kappa_1 |\Phi_{ \repb{3}\,,1} |^2  + \kappa_2 |\Phi_{ \repb{3}\,,2} |^2  \Big) |  \Phi_{ \repb{3}}^\prime |^2 \non
&+& \kappa_3 (\Phi_{ \repb{3}\,, 1 }^\dag \Phi_{ \repb{3}}^\prime ) (\Phi_{ \repb{3}}^{\prime\,\dag} \Phi_{ \repb{3}\,, 1 } ) + \kappa_4  (\Phi_{ \repb{3}\,, 2 }^\dag \Phi_{ \repb{3}}^\prime ) (\Phi_{ \repb{3}}^{\prime\,\dag} \Phi_{ \repb{3}\,, 2 } ) \non
& +& \Big(  \nu \epsilon_{IJK} (\Phi_{ \repb{3}\,,1})_I ( \Phi_{ \repb{3}\,, 2} )_J  (\Phi_{ \repb{3}}^\prime)_K  + H.c. \Big)  \,. \label{eq:2HTM_potential03}
\eeqn
\eeqs
The last $\nu$-term in Eq.~\eqref{eq:2HTM_potential03} is inevitable by both the gauge symmetry and the emergent global ${\rm SU}(2)_F$ symmetry in Eq.~\eqref{eq:SU6_flavor}, with $I\,,J\,,K=1\,,2\,,3$ being the ${\rm SU}(3)_W$ anti-fundamental indices.
\bigskip
%
%




We denote the Higgs fields under the ${\rm SU}(3)_W \otimes {\rm U}(1)_{X}$ representations as follows
\beqn
&&  \Phi_{ \repb{3}\,, \rho }  = \frac{1}{ \sqrt{2}} \left( \ba{c}  \sqrt{2} \phi_\rho^- \\    \phi_\rho - i \eta_\rho   \\    h_\rho - i\pi_\rho \ea  \right)\,,\quad  \Phi_{ \repb{3} }^\prime  = \frac{1}{ \sqrt{2}} \left( \ba{c}  h_u + i \pi_u \\   \sqrt{2} \chi^+  \\  \sqrt{2} \chi^{\prime\,+}   \ea  \right) \,,
\eeqn
where the electric charges are given according to Eq.~\eqref{eq:Qcharge_331}.
According to the VEV assignment in Eq.~\eqref{eq:331_VEVnaive}, we expect the non-vanishing Higgs VEVs of
\beqn\label{eq:331_VEV01}
&& \langle h_1 \rangle= V_1 =  V_{331} c_{\tilde \beta} \,,\quad \langle h_2 \rangle= V_2 =  V_{331} s_{\tilde \beta} \,,\quad \langle h_u \rangle= v_u  \,,
\eeqn
with 
\beqn
t_{\tilde \beta }&\equiv& \frac{V_2}{ V_1}
\eeqn
parametrizing the ratio between two $331$ symmetry breaking VEVs.
Accordingly, the minimization of the Higgs potential in Eqs.~\eqref{eqs:2HTM_potential} leads to the following conditions
\beqs\label{eqs:2HTM_min}
\beqn
\frac{ \partial V}{ \partial V_1} =0&\Rightarrow& m_{11}^2= m_{12}^2 t_{\tilde \beta} - \frac{ \lambda_1}{2} V_1^2 - \frac{ \lambda_{3} + \lambda_4 + \lambda_5 }{2}  V_2^2 - \frac{ \kappa_{1}}{2} v_u^2   \,,\\[1mm]
\frac{ \partial V}{ \partial V_2} =0&\Rightarrow& m_{22}^2= \frac{ m_{12}^2 }{ t_{\tilde \beta} } - \frac{ \lambda_2}{2} V_2^2 - \frac{ \lambda_{3} + \lambda_4 + \lambda_5 }{2} V_1^2 - \frac{ \kappa_{2}}{2} v_u^2 \,,\\[1mm]
\frac{ \partial V}{ \partial v_u} =0&\Rightarrow& m^2= -\lambda v_u^2 - \hf ( \kappa_{1} V_1^2 + \kappa_{2} V_2^2 )  \,.
\eeqn
\eeqs
Note that the $\nu$-term that mixes the $\Phi_{ \repb{3}\,,\rho }$ and the $\Phi_{ \repb{3} }^\prime$ does not enter the minimization condition with the VEV assignment in Eq.~\eqref{eq:331_VEVnaive}, while this term will be important in generating the fermion masses.
Correspondingly, this leads to an unwanted tadpole term of $- \frac{1}{2 \sqrt{2} } \nu ( \phi_1 V_2 - \phi_2 V_1) v_u $ in the Higgs potential.
\bigskip

To resolve the tadpole problem, the only way is to develop EWSB VEVs of
\beqn\label{eq:331_VEV02}
&&\langle \phi_1\rangle = u_1=v_\phi c_{\beta^\prime} \,,\quad \langle \phi_2\rangle = u_2 = v_\phi s_{\beta^\prime} \,,
\eeqn
presumably with $v_\phi \sim \Oc(v_{\rm EW} )$ as was considered in Refs.~\cite{Boucenna:2014ela,Boucenna:2014dia,Boucenna:2015zwa,Deppisch:2016jzl}.
The Nambu-Goldstone boson (NGB) of $\xi^0$ can be obtained from the following derivative terms
\beqn
&\sim& \frac{i }{ 2\sqrt{2} } g_{3L} ( N_\mu - \bar N_\mu ) \partial^\mu \Big[  ( u_1 h_1 + u_2 h_2 ) - ( V_1 \phi_1 + V_2 \phi_2 )  \Big]  \non
&\Rightarrow& \xi^0= c_\theta ( c_{\tilde \beta} \phi_1 + s_{\tilde \beta} \phi_2 ) - s_\theta ( c_{\beta^\prime}  h_1 + s_{\beta^\prime } h_2 ) \,,
\eeqn
with 
\beqn
t_\theta &\equiv& \frac{ v_\phi }{ V_{331} }
\eeqn
parametrizing the ratio between two symmetry-breaking scales in each ${\rm SU}(3)_W$ anti-fundamental Higgs of $\Phi_{ \repb{3}\,,\rho }$.
With a natural assumption of the following VEV orthogonal relation of
%
\beqn\label{eq:betas}
&& \sum_\rho u_\rho V_\rho =0  \Rightarrow \beta^\prime = \tilde \beta - \frac{\pi}{2} \,,
\eeqn
the potential mixing between the $(W_\mu^\pm\,, C_\mu^\pm)$, as well as the $( N_\mu\,, \bar N_\mu \,, Z_\mu^\prime)$, can be avoided.
This can be confirmed with the explicit gauge fields in terms of a $3\times 3$ matrix given in Eq.~\eqref{eq:331_connection}.
Thus, the VEVs in Eq.~\eqref{eq:331_VEV02} become
\beqn\label{eq:331_VEV03}
&&\langle \phi_1\rangle = u_1=v_\phi s_{\tilde \beta } \,,\quad \langle \phi_2\rangle = u_2 = - v_\phi c_{\tilde \beta } \,.
\eeqn
\bigskip

The minimization conditions of the Higgs potential in Eqs.~\eqref{eqs:2HTM_min} are modified to be
\beqs\label{eqs:2HTM_min_mod}
\beqn
\frac{\partial V}{\partial V_1} =0 &\Rightarrow& m_{11}^2 = m_{12}^2 t_{\tilde \beta} - \frac{\lambda_1}{2}(  V_1^2 +  u_1^2 )  - \frac{ \lambda_{3}}{2} ( V_2^2 +  u_2^2 ) \non
&&- \frac{ \lambda_4 + \lambda_5 }{2}  ( u_1 u_2 + V_1 V_2 ) t_{\tilde \beta} -  \frac{ \kappa_{1}}{2}   v_u^2 + \frac{ \nu u_2 v_u}{  \sqrt{2} V_1}  \,,\label{eq:2HTM_min_mod01} \\[1mm]
\frac{\partial V}{\partial V_2} =0 &\Rightarrow& m_{22}^2 = \frac{ m_{12}^2}{ t_{\tilde\beta} } - \frac{ \lambda_2}{2}( V_2^2 + u_2^2 ) -\frac{ \lambda_{3}}{2} ( V_1^2 +  u_1^2 ) \non
&&- \frac{ \lambda_4 + \lambda_5 }{2}  ( u_1 u_2 + V_1 V_2 ) \frac{1 }{t_{\tilde \beta}} -  \frac{ \kappa_{2}}{2} v_u^2 - \frac{ \nu u_1 v_u}{ \sqrt{2} V_2} \,, \label{eq:2HTM_min_mod02} \\[1mm]
\frac{\partial V}{\partial v_u} =0 &\Rightarrow& - m^2 = \lambda v_u^2 +  \frac{\kappa_{1} }{2}(V_1^2 + u_1^2 ) + \frac{\kappa_{2}}{2} ( V_2^2 + u_2^2 ) \non
&& + \frac{ \nu }{ \sqrt{2}  v_u} ( u_1 V_2  - u_2 V_1 )  \,, \label{eq:2HTM_min_mod03}\\[1mm]
\frac{\partial V}{\partial u_1} =0 &\Rightarrow&  m_{11}^2 =  - \frac{m_{12}^2}{ t_{\tilde \beta} }  - \frac{\lambda_1 }{2} (  V_1^2 +  u_1^2 ) - \frac{ \lambda_{3}}{2} ( V_2^2 + u_2^2 ) \non
&&+  \frac{ \lambda_4 + \lambda_5 }{2}  ( u_1 u_2 + V_1 V_2 ) \frac{1 }{t_{\tilde \beta}} -  \frac{ \kappa_{1}}{2}  v_u^2 - \frac{ \nu  V_2 v_u}{ \sqrt{2} u_1} \,,\label{eq:2HTM_min_mod04} \\[1mm]
\frac{\partial V}{\partial u_2} =0 &\Rightarrow&  m_{22}^2 =  -m_{12}^2 t_{ \tilde \beta }  - \frac{ \lambda_2}{2}( V_2^2 + u_2^2 ) -\frac{ \lambda_{3}}{2} ( V_1^2 + u_1^2 ) \non
&&+  \frac{ \lambda_4 + \lambda_5 }{2}  ( u_1 u_2 + V_1 V_2 ) t_{\tilde \beta }  - \frac{ \kappa_{2}}{2}  v_u^2 + \frac{ \nu  V_1 v_u}{ \sqrt{2} u_2}  \,.\label{eq:2HTM_min_mod05}
\eeqn
\eeqs
By equating Eqs.~\eqref{eq:2HTM_min_mod01} with \eqref{eq:2HTM_min_mod04}, and Eqs.~\eqref{eq:2HTM_min_mod02} with \eqref{eq:2HTM_min_mod05}, we have a constraint of 
\beqn\label{eq:331_constraint}
&&  t_\theta^3 -\frac{ \sqrt{2} \nu v_{u}  }{(\lambda_{4}+\lambda_{5}) V_{331}^2 } t_{\theta}^2 + \big(  \frac{ 2 m_{12}^2 }{s_{ \tilde \beta} c_{ \tilde \beta} (\lambda_{4}+\lambda_{5}) V_{331}^2 } - 1 \big) t_\theta + \frac{ \sqrt{2} \nu v_{u}  }{(\lambda_{4}+\lambda_{5}) V_{331}^2 }=0 \,,
\eeqn
with the relation in Eq.~\eqref{eq:betas}.
Consider the scale hierarchy of $m_{12}^2 \sim \Oc(V_{331}^2)$, $\nu \sim v_u \sim \Oc(v_{\rm EW})$.
%
%
%
%
It is straightforward to expect $t_\theta \sim \Oc( (v_{\rm EW}/ V_{331} )^2 )$ from Eq.~\eqref{eq:331_constraint}.
Thus, a natural scale of $v_\phi$ can be further suppressed from the EW scale of $v_{\rm EW}$, such as $\sim \Oc(1)\,\GeV$.
This was not noted in the previous Refs.~\cite{Boucenna:2014ela,Boucenna:2014dia,Boucenna:2015zwa,Deppisch:2016jzl} with the similar VEV assignment.
Here and after, we will consider the following parameter inputs~\footnote{A negative $\nu$ is chosen so that the physical Higgs boson mass squares are positive in the current context. This may not be necessary with additional terms included in the Higgs potential.}
\beqn\label{eq:BM_input}
&& \nu \sim -\Oc(100)\,{\rm GeV}\,, \quad v_u \approx 246\,{\rm GeV}\,,\quad v_\phi \sim \Oc(1)\,{\rm GeV} \,,\quad V_{331} \sim \Oc(10)\,{\rm TeV} \non
&& t_{\tilde \beta }\sim \Oc(1) \,, \quad m_{12}^2 \sim  \Oc(100)\,{\rm TeV}^2 \,,
\eeqn
instead of performing the detailed parameter scans.
The choice of $V_{331}\sim \Oc(10)\,{\rm TeV}$ will become clear from the $(b\,,\tau)$ Yukawa couplings with the $125\,\GeV$ SM-like Higgs boson.
Notice that in the decoupling limit of $m_{12}\sim V_{331}\to \infty$, one naturally has $t_\theta\to 0$ and $v_\phi\to 0$ from Eq.~\eqref{eq:331_constraint}.
\bigskip

\subsection{The charged and CP-odd Higgs bosons}

The charged Higgs bosons of $\Phi_{\pm}=(  \phi_1^\pm \,,  \phi_2^\pm \,,  \chi^\pm \,,  \chi^{\prime\, \pm})$ form the mass squared matrix of
%
%
\beqn\label{eq:331_Charged}
( \Mc_{\pm }^2)_{ 4\times 4}  &=& \left( \begin{array}{cccc}
 M_{\phi_1^+ \phi_1^-}^2 &  M_{\phi_1^+ \phi_2^- }^2   &  M_{\phi_1^+ \chi^- }^2  &  M_{\phi_1^+ \chi^{\prime\,-} }^2   \\[2mm]
M_{\phi_1^- \phi_2^+ }^2 &  M_{\phi_2^+ \phi_2^- }^2   &  M_{\phi_2^+ \chi^- }^2  &  M_{\phi_2^+ \chi^{\prime\, -}  }^2   \\[2mm]
M_{\phi_1^- \chi^+ }^2  &  M_{\phi_2^- \chi^+ }^2   & M_{\chi^+ \chi^- }^2 &  M_{\chi^+ \chi^{\prime\,-} }^2  \\[2mm]
 M_{\phi_1^- \chi^{\prime\,+} }^2  &  M_{\phi_2^- \chi^{\prime\, +} }^2   & M_{\chi^- \chi^{\prime\, +} }^2 &  M_{\chi^{\prime\,+}  \chi^{\prime\,-} }^2  \\
 \end{array}  \right) 
\sim \left( \begin{array}{cccc}
1  &  1  & \epsilon  & \epsilon  \\
 1 &  1  & \epsilon  &  \epsilon  \\
 \epsilon & \epsilon &  \epsilon^2 &  \epsilon^2  \\
 \epsilon & \epsilon & \epsilon^2  & 1   \\
\end{array}  \right) V_{331}^2\,,
\eeqn
%
%
with a small parameter given by $\epsilon\equiv \frac{v_u}{V_{331} }\sim \frac{v_\phi }{ \nu } \sim \Oc(0.01)$.
The orthogonal transformations to gauge eigenstates of $\Phi_{\pm}$ are expressed as follows
\beqn\label{eq:331Charged_transf}
 \left( \ba{c}   G^\pm \\  G^{\prime\, \pm}  \\  H_1^\pm \\ H_2^\pm  \ea \right)  &=& \widetilde V_{\pm} \cdot \left( \ba{c}   \phi_1^\pm \\  \phi_2^\pm \\ \chi^\pm \\  \chi^{\prime\, \pm}  \ea \right)  \,, \non
  \widetilde V_{\pm} &=& \left( 
 \ba{cccc}   
 c_{\tau_1} & 0  &  0 & -s_{\tau_1}    \\  
 0 & 1  &  0 &  0   \\  
 0  & 0 & 1 &   0  \\   
 s_{\tau_1}  & 0  & 0  &  c_{\tau_1}   \\  
  \ea \right) \cdot  
  \left( 
 \ba{cccc}   
 1 & 0  &  0 & 0    \\  
 0 & c_{\tau_2}  & -s_{\tau_2} &  0   \\  
 0  & s_{\tau_2} & c_{\tau_2} &   0  \\   
 0  & 0  & 0  &  1   \\  
  \ea \right) \cdot 
   \left( 
 \ba{cccc}   
 c_{\tilde \beta} & s_{\tilde \beta}  &  0 & 0    \\  
 - s_{\tilde \beta} & c_{\tilde \beta}  &0 &  0   \\  
 0  & 0 & 1 &   0  \\   
 0  & 0  & 0  &  1   \\  
  \ea \right)  \non
  &=& \left( 
 \ba{cccc}   
 c_{\tau_1} c_{\tilde \beta} &  c_{\tau_1} s_{\tilde \beta}  &  0 & -s_{\tau_1}    \\  
  -c_{\tau_2} s_{\tilde \beta}  & c_{\tau_2} c_{\tilde \beta}   &  -s_{\tau_2} &  0   \\  
 -s_{\tau_2} s_{\tilde \beta}  & s_{\tau_2} c_{\tilde \beta}  &  c_{\tau_2} &   0  \\   
 s_{\tau_1} c_{\tilde \beta}  & s_{\tau_1} s_{\tilde \beta}   & 0  &  c_{\tau_1}   \\  
  \ea \right)  \,,
\eeqn
with $t_{\tau_1} \equiv \frac{v_u}{ V_{331}} $ and $t_{\tau_2} \equiv \frac{v_\phi }{v_u} $.
Two corresponding non-zero eigenvalues for two charged Higgs bosons of $H_{1\,,2}^\pm$ are given by
\beqn\label{eq:331_ChargedHmass}
&&  M_{H_1^\pm }^2 =\frac{\Bc }{ 16 V_{331} v_u v_\phi} \Big( 1 - \sqrt{1 + \frac{\Ac}{ \Bc^2}} \Big) \,, \quad M_{H_2^\pm }^2 = \frac{ \Bc }{ 16 V_{331} v_u v_\phi} \Big( 1 + \sqrt{1 + \frac{\Ac }{\Bc^2}} \Big) \,, \non
&& \Ac = 32 \sqrt{2} \nu v_{u} v_{\phi} (v_{u}^2+v_{\phi}^2)(\kappa_{3}+\kappa_{4}+(\kappa_{3}-\kappa_{4}) c_{2 \tilde \beta}) V_{331}^5- 64 v_{\phi}^2(v_{u}^2+v_{\phi}^2)(2\nu^2+\kappa_{3} \kappa_{4} v_{u}^2) V_{331}^4 \non
&+&  32 \sqrt{2} \nu v_{u} v_{\phi} (v_{u}^2+v_{\phi}^2)((v_{u}^2+v_{\phi}^2)(\kappa_{3}+\kappa_{4})+(v_{u}^2-v_{\phi}^2)(\kappa_{3}-\kappa_{4}) c_{2 \tilde \beta}) V_{331}^3 \non
&-&64 v_{u}^2 v_{\phi}^2 (v_{u}^2+v_{\phi}^2)(2 \nu^2+\kappa_{3} \kappa_{4} v_{u}^2)  V_{331}^2 -  32 \sqrt{2} \nu v_{u}^3 v_{\phi}^3 (v_{u}^2 + v_{\phi}^2)((\kappa_{3}-\kappa_{4})c_{2 \tilde \beta}-\kappa_{3}-\kappa_{4})  V_{331} \non
&\sim& \nu v_\phi v_u^3 V_{331}^5 \,, \non
&& \Bc = 2 V_{331} v_{u} v_{\phi} \Big[ c_{2 \tilde \beta} (\kappa_{3}-\kappa_{4}) (V_{331}^2-v_{\phi}^2)+ (\kappa_{3}+\kappa_{4}) (V_{331}^2+v_{\phi}^2) \Big] \non
&-&8 \sqrt{2} \nu V_{331}^2 v_{\phi}^2  - 4 \sqrt{2} \nu v_{u}^2 (V_{331}^2+v_{\phi}^2)+4 V_{331} v_{u}^3 v_{\phi} (\kappa_{3}+\kappa_{4})  \non
&\sim& v_u v_\phi V_{331}^3 \,.
\eeqn
%
%
A simple expansion of Eq.~\eqref{eq:331_ChargedHmass} in terms of the mass hierarchy assumed in Eq.~\eqref{eq:BM_input} leads to the approximated mass scales of
\beqn\label{eq:331_ChargedHmass_approx}
&& M_{H_1^\pm }^2  \sim \Oc( V_{331}^2 )  \,,\quad  M_{H_2^\pm }^2 \sim \Oc( V_{331}^2 )\,.
\eeqn
\bigskip

The CP-odd Higgs bosons form the mass squared matrix of
\beqn\label{eq:331_CPodd}
( \Mc_{0^- }^2)_{ 5\times 5}  &=&  \left(
\begin{array}{ccccc}
M_{\pi_u  \pi_u}^2 & M_{\pi_u \eta_1}^2 & M_{\pi_u \eta_2}^2 & M_{\pi_u \pi_1}^2 & M_{\pi_u \pi_2}^2 \\
M_{\pi_u \eta_1}^2 & M_{\eta_1 \eta_1}^2 & M_{\eta_1 \eta_2}^2 & M_{\pi_1 \eta_1}^2 & M_{\pi_2 \eta_1}^2 \\
M_{\pi_u \eta_2}^2 & M_{\eta_1 \eta_2}^2 & M_{\eta_2  \eta_2}^2 & M_{\pi_1 \eta_2}^2 & M_{\pi_2 \eta_2}^2 \\
M_{\pi_u \pi_1}^2 & M_{\pi_1 \eta_1 }^2 & M_{\pi_1 \eta_2}^2 & M_{\pi_1 \pi_1}^2 & M_{\pi_1 \pi_2}^2  \\
M_{\pi_u \pi_2}^2 & M_{\pi_2 \eta_1}^2 & M_{\pi_2 \eta_2}^2 & M_{\pi_1 \pi_2}^2 & M_{\pi_2 \pi_2}^2  \\
\end{array} \right) \non
&\sim& \left(
\begin{array}{ccccc}
\epsilon^2 & \epsilon  & \epsilon  &  \epsilon^3 &  \epsilon^3  \\
 \epsilon & 1  & 1  & \epsilon^2  &  \epsilon^2  \\
\epsilon  & 1  & 1  & \epsilon^2  &  \epsilon^2  \\
 \epsilon^3 & \epsilon^2  & \epsilon^2  & 1  & 1   \\
 \epsilon^3 &  \epsilon^2 & \epsilon^2  & 1  &  1  \\
\end{array} \right) V_{331}^2\,,
\eeqn
in the basis of  $\Phi_{0^-}=(\pi_u\,,  \eta_{1\,,2}\,,  \pi_{1\,,2})$.
We find three zero eigenvalues corresponding to three massless NGBs with the constraint in Eq.~\eqref{eq:331_constraint}.
The orthogonal transformations to $\Phi_{0^-}$ are expressed as follows
\beqn\label{eq:331CPodd_transf}
 \left( \ba{c}   G^0 \\  G^{0\,\prime}  \\  G^{0\,\prime\prime } \\  A^{0} \\ A^{0\,\prime}  \ea \right)  &=& \widetilde V_{0^-} \cdot \left( \ba{c}   \pi_u \\  \eta_1  \\ \eta_2  \\ \pi_1 \\ \pi_2 \\  \ea \right)  \,, \quad \widetilde V_{0^-} = \left( \ba{ccccc}   
 0 & c_\theta c_{\tilde \beta} & c_\theta s_{\tilde \beta}  & s_\theta s_{\tilde \beta}  & -s_\theta c_{\tilde \beta}  \\  
 0 & - s_\theta s_{\tilde \beta} & s_\theta c_{\tilde \beta}  & c_\theta c_{\tilde \beta}  & c_\theta s_{\tilde \beta}  \\  
c_\tau  & s_\tau c_\theta s_{\tilde \beta} &  - s_\tau c_\theta c_{\tilde \beta} &  s_\tau s_\theta c_{\tilde \beta} & s_\tau s_\theta s_{\tilde \beta}  \\  
 -s_\tau & c_\tau c_\theta s_{\tilde \beta} & -c_\tau c_\theta c_{\tilde \beta}  & c_\tau s_\theta c_{\tilde \beta}  & c_\tau s_\theta s_{\tilde \beta}  \\  
 0 & s_\theta c_{\tilde \beta} &  s_\theta s_{\tilde \beta}  & - c_\theta s_{\tilde \beta}  & c_\theta c_{\tilde \beta}  \\  
  \ea \right)  \,,
\eeqn
with $t_\tau \equiv \frac{ v_\phi V_{331} }{ \sqrt{ V_{331}^2 + v_\phi^2 }  v_u } \approx \frac{v_\phi }{v_u }$.
Two corresponding non-zero eigenvalues for two CP-odd Higgs bosons are given by
\beqn\label{eq:331_CPoddmass}
&& M_{A^0 }^2 = \frac{-\sqrt{2} \nu (V_{331}^2 + v_\phi^2 ) v_u^2+(\lambda_{4} - \lambda_5 ) V_{331} v_\phi ( V_{331}^2  + v_{\phi}^2 ) v_u   }{2 V_{331} v_{\phi} v_u} \sim \Oc( V_{331}^2 )  \,,\non
&& M_{A^{0\,\prime} }^2 = - \nu \frac{ V_{331}^2 ( v_u^2 + v_\phi^2 ) +  v_{\phi}^2 v_u^2}{ \sqrt{2} V_{331} v_{\phi} v_u} \sim \Oc( V_{331}^2 ) \,.
\eeqn
Hence, we do not expect the discovery of these two CP-odd Higgs bosons at the current LHC direct searches.
\bigskip

\subsection{The CP-even Higgs bosons}

There are five CP-even Higgs fields of $(h_u\,, \phi_{1\,,2}\,,  h_{1\,,2})$ in the gauge eigenbasis, and one of their linear combination will be identified as the NGB. 
Their masses and mixings play the key role in generating the bottom quark and tau lepton masses, as well as determining their Yukawa couplings with the SM-like Higgs boson.
It takes two steps to obtain their mass eigenstates.
To see that, we first perform the orthonormal transformations to $(  \phi_{1\,,2}\,,  h_{1\,,2} )$ as follows
\beqn\label{eq:331CPeven_transf01}
\left( \ba{c}  \xi^0 \\  \phi^0 \\  h_1^\prime  \\ h_2^\prime  \\  \ea \right)  &=&\widetilde V_{0^+} \cdot \left( \ba{c}   \phi_1 \\  \phi_2  \\ h_1 \\ h_2 \\  \ea \right)  \,, \quad  \widetilde V_{0^+}= \small{  \left(
\begin{array}{cccc} 
 c_\theta c_{\tilde \beta}  &  c_\theta s_{\tilde \beta}  & - s_\theta s_{\tilde \beta}  & s_\theta c_{\tilde \beta}  \\
 - c_\theta s_{\tilde \beta}  &  c_\theta c_{\tilde \beta}  & - s_\theta c_{\tilde \beta}  &  - s_\theta s_{\tilde \beta}  \\
 s_\theta c_{\tilde \beta} & s_\theta s_{\tilde \beta}  & c_\theta s_{\tilde \beta}  & - c_\theta c_{\tilde \beta} \\
- s_\theta s_{\tilde \beta}  & s_\theta c_{\tilde \beta}  &  c_\theta c_{\tilde \beta}  & c_\theta s_{\tilde \beta}  \\   \end{array} \right) } \,,
\eeqn
with $\xi^0$ being the massless NGB.
Under the basis of $(h_u\,, \phi^0\,, h_{1\,,2}^\prime )$, the remaining four CP-even Higgs fields form the mass squared matrix and can be expanded as follows
\beqn\label{eq:331_CPevenHiggs}
( \Mc_{0^+ }^2)_{ 4\times 4} &=&  \left( 
\begin{array}{cccc}  
  M_{h_u h_u}^2  &  M_{ h_u \phi^0 }^2  & M_{h_u h_1^\prime}^2   &  M_{h_u h_2^\prime }^2   \\[2mm]
  M_{h_u \phi^0 }^2  &   M_{\phi^0 \phi^0 }^2 & M_{\phi^0 h_1^\prime}^2  &  M_{\phi^0 h_2^\prime }^2   \\[2mm]
  M_{h_u h_1^\prime}^2   & M_{\phi^0 h_1^\prime }^2   & M_{ h_1^\prime h_1^\prime }^2  &   M_{ h_1^\prime h_2^\prime }^2   \\[2mm]
 M_{h_u h_2^\prime }^2   &  M_{\phi^0 h_2^\prime }^2  &  M_{ h_1^\prime h_2^\prime  }^2  & M_{ h_2^\prime h_2^\prime }^2    \\   \end{array} \right) \sim \left(
\begin{array}{cccc}
 \epsilon^2  & \epsilon &  \epsilon & \epsilon  \\
 \epsilon   & 1 & \epsilon^2 & \epsilon^2 \\
\epsilon  & \epsilon^2 &  1 &  1
   \\
 \epsilon & \epsilon^2 & 1 &  1 \\
\end{array}
\right) V_{331}^2 \non
&=& ( \Mc_{0^+ }^2)^{(0)} + ( \Mc_{0^+ }^2)^{(1)} + ( \Mc_{0^+ }^2)^{(2)} \,,
\eeqn
with a small parameter given by $\epsilon\equiv \frac{v_u}{V_{331} }\sim \Oc(0.01)$.
The further diagonalization of Eq.~\eqref{eq:331_CPevenHiggs} transforms into mass eigenstates of $(H_u\,, H_\phi \,, H_1 \,, H_2)$ such that
\beqn\label{eq:331_CPeven_mix}
&& \left( \ba{c}  H_u \\ H_\phi \\ H_1 \\ H_2 \ea \right) = V_{0^+}\cdot \left( \ba{c}  h_u \\  \phi^0  \\ h_1^\prime  \\ h_2^\prime  \\  \ea \right) \,, \quad V_{0^+} ( \Mc_{0^+ }^2) V_{0^+}^T = {\rm diag} ( M_{H_u}^2\,, M_{H_\phi}^2\,, M_{H_1}^2 \,, M_{H_2}^2 )\,.
\eeqn 
Among them, $H_u$ is the lightest CP-even Higgs boson with mass of $125$ GeV, while others have masses of $\sim \Oc(V_{331})$.
\bigskip

To have positive definite eigenvalues for CP-even Higgs boson mass squares in Eq.~\eqref{eq:331_CPevenHiggs}, one cannot have the $\nu$ parameter as large as $V_{331}$.
That is why we chose $\nu\sim \Oc(100)\,{\rm GeV}$ in Eq.~\eqref{eq:BM_input}.
However, a $\nu$-problem emerges, namely, why is a mass parameter from a $331$-invariant Higgs potential takes a value comparable to the EW scale.
This problem is analogous to the well-known $\mu$-problem in the minimal supersymmetric Standard Model (MSSM)~\cite{Martin:1997ns}.
One can thus expect this $\nu$-term to originate from some non-renormalizable terms in the realistic non-minimal GUTs with $n_g=3$~\footnote{In the MSSM case, this was known as the Kim-Nilles mechanism for the $\mu$-problem~\cite{Kim:1983dt,Murayama:1992dj}.}.
This type of terms are inevitable due to the gravitational effects that break the global ${\rm U}(1)$ symmetry explicitly~\footnote{In the axion models, this effect leads to what is known as the PQ quality problem~\cite{Barr:1992qq,Kamionkowski:1992mf,Holman:1992us}.}.
Taking the ${\rm SU}(8)$ GUT as an example again, one such possible $d=5$ non-renormalizable term is expected to be 
\beqn
{\rm SU}(8)~&:&~ \frac{1}{ M_{\rm pl} }  \epsilon_{\rho_1 ... \rho_4 } \repb{8_H}^{\rho_1} ... \repb{8_H}^{\rho_4 } \cdot \rep{70_H}  \non
&\supset& \frac{1}{ M_{\rm pl} }  \epsilon_{\rho_1 ... \rho_4 } \, ( \repb{4}\,, \rep{1}\,, +\frac{1}{4} )_{ \mathbf{H}}^{\rho_1} \otimes ( \rep{1}\,, \repb{4}\,, -\frac{1}{4} )_{ \mathbf{H}}^{\rho_2} \otimes ( \rep{1}\,, \repb{4}\,, -\frac{1}{4} )_{ \mathbf{H}}^{\rho_3} \otimes ( \rep{1}\,, \repb{4}\,, -\frac{1}{4} )_{ \mathbf{H}}^{\rho_4}  \otimes  ( \rep{4}\,, \repb{4}\,, +\frac{1}{2} )_{ \mathbf{H}} \non 
&\supset&  \frac{ V_{441} }{ M_{\rm pl} }  \epsilon_{\rho_2 ... \rho_4 } \, ( \rep{1}\,, \repb{4}\,, -\frac{1}{4} )_{ \mathbf{H}}^{\rho_2} \otimes ( \rep{1}\,, \repb{4}\,, -\frac{1}{4} )_{ \mathbf{H}}^{\rho_3} \otimes  ( \rep{1}\,, \repb{4}\,, -\frac{1}{4} )_{ \mathbf{H}}^{\rho_4}  \otimes ( \rep{1}\,, \repb{4}\,, +\frac{3}{4} )_{ \mathbf{H}}   \non
&\sim& \frac{ V_{441} V_{341} }{ M_{\rm pl} }   \epsilon_{ \rho_3  \rho_4} \, ( \rep{1}\,, \repb{3}\,, -\frac{1}{3} )_{ \mathbf{H}}^{\rho_3} \otimes ( \rep{1}\,, \repb{3}\,, -\frac{1}{3} )_{ \mathbf{H}}^{\rho_4} \otimes  ( \rep{1}\,, \repb{3}\,, +\frac{2}{3} )_{ \mathbf{H}}^{\prime \prime \prime}  \,.
\eeqn
Here, the decompositions are achieved according to Eqs.~\eqref{eqs:SU8_HiggsDecomp}.
Obviously, this non-renormalizable term induced by the gravitational effect reproduces what we considered as the $\nu$-term in Eq.~\eqref{eq:2HTM_potential03}.
Thus, the value of $\nu\sim \Oc(100)\,{\rm GeV}$ in Eq.~\eqref{eq:BM_input} can be naturally realized with $V_{441}\sim \Oc(10^{12})\,{\rm GeV}$ and $V_{341}\sim \Oc(10^{9})\,{\rm GeV}$.
\bigskip

With the hierarchies of mass parameter in Eq.~\eqref{eq:BM_input}, the diagonalization of mass matrix in Eq.~\eqref{eq:331_CPevenHiggs} can be achieved in terms of perturbation.
Hence, we express the mixing matrix in Eq.~\eqref{eq:331_CPeven_mix} as
\beqn\label{eq:331_CPeven_mix_expr}
V_{0^+}&=& \widetilde U  \widetilde U^{(0) } \,.
\eeqn
At the leading order, it is straightforward to diagonalize the $( \Mc_{0^+ }^2)^{(0)}$ by an orthogonal matrix of
\beqn\label{eq:331_CPeven_mix0}
 \widetilde U^{(0)}&=& \left(
\begin{array}{ccc} 
\mathbb{I}_{2\times 2} &  &  \\ 
  & c_\alpha & - s_\alpha  \\  
  & s_\alpha & c_\alpha  \\  \end{array} \right) \,,
\eeqn
into
\beqn
&&  \widetilde U^{(0)} \cdot ( \Mc_{0^+ }^2)^{(0)} \cdot  \widetilde U^{(0)\, T} = {\rm diag}(  0\,, M_{\phi^0 \phi^0 }^2 \,,  \widetilde M_{h_1^\prime  h_1^\prime}^{ 2} \,, \widetilde M_{h_2^\prime  h_2^\prime}^{ 2} )\,,
\eeqn
where $  t_{\alpha}=[ M_{h_1^\prime h_1^\prime}^2- M_{h_2^\prime h_2^\prime}^2 -\sqrt{( M_{h_1^\prime h_1^\prime}^2 -  M_{h_2^\prime h_2^\prime}^2  )^2 + 4 M_{h_1^\prime h_2^\prime}^4}]/(2 M_{h_1^\prime h_2^\prime}^2) $. The mixing matrix of $\widetilde U$ for the higher-order terms can be expanded up to $\Oc(\epsilon^2)$ as
\beqn
\widetilde U &=& \mathbb{I}_{4\times 4} +  \widetilde U^{(1)} +  \widetilde U^{ (2)} + \Oc(\epsilon^3) \,,
\eeqn
with $\widetilde U^{(1)} \sim \Oc(\epsilon)$ and $\widetilde U^{(2)} \sim \Oc(\epsilon^2)$. Similarly, $V_{0^+}$ can also be expanded as:
\beqn\label{eq:331_CPeven_expand}
V_{0^+} &=& \widetilde U^{(0)} +  V_{0^+}^{(1)} + V_{0^+}^{(2)}  + \Oc(\epsilon^3) \,.
\eeqn
For our later usage, we find that the $ V_{0^+}^{(1)} $ is expressed as follows
\beqn\label{eq:331_CPeven_mix1}
V_{0^+}^{(1)}&=& \small{ \left(
\begin{array}{cccc} 
  & - \frac{ M_{ h_u \phi^0}^2 }{M_{\phi^0 \phi^0 }^2 }  & - \frac{ M_{ h_u h_1^\prime }^{ 2} }{M_{ h_1^\prime h_1^\prime}^{2 }}  & - \frac{ M_{ h_u  h_2^\prime}^{2 } }{M_{ h_2^\prime  h_2^\prime}^{ 2} } \\ 
 \frac{ M_{ h_u \phi^0 }^2 }{M_{\phi^0 \phi^0 }^2 } &  &   &   \\[2mm]  
\frac{ M_{ h_u  h_1^\prime}^{2 } }{M_{ h_1^\prime h_1^\prime }^{ 2} }  &  &  &   \\[2mm] 
 \frac{ M_{ h_u  h_2^\prime}^{2 } }{M_{ h_2^\prime h_2^\prime }^{ 2} }  &  &  &   \\  \end{array} \right) \sim  \left(
\begin{array}{cccc} 
  & - \epsilon  & - \epsilon   & - \epsilon  \\ 
\epsilon  &  &   &   \\  
\epsilon  &  &  &   \\ 
\epsilon  &  &  &   \\  \end{array} \right)  } \,,
\eeqn
with $(0 \,, 0\,, \widetilde M_{h_u h_1^\prime}^{2 }\,, \widetilde M_{h_u h_2^\prime }^{ 2} ) = \widetilde U^{(0)} \cdot ( 0 \,, 0\,, M_{ h_u h_1^\prime }^2 \,, M_{h_u h_2^\prime }^2 )  $.
By using the perturbation expansion in Eq.~\eqref{eq:331_CPeven_expand}, we find the SM-like CP-even Higgs boson mass of
\beqn
M_{H_u}^2&=& (2 \lambda - \frac{ \nu v_\phi V_{331} }{ \sqrt{2} v_u^3 } ) v_u^2 - [ (V_{0^+}^{(1)} )_{12}^2 M_{\phi^0  \phi^0 }^2 ] - [ (V_{0^+}^{(1)} )_{13}^2 M_{h_1^\prime h_1^\prime}^2 ]  -[ (V_{0^+}^{(1)} )_{14}^2 M_{h_2^\prime h_2^\prime}^2 ]\,.
\eeqn
Since the mixing elements are $(V_{0^+}^{(1)})_{ij}\sim \Oc(\epsilon)$, all terms here are of the EW scales.
\bigskip

\subsection{Summary of the Higgs spectrum}
\label{sec:331Hspec}

\begin{table}[htp]
\begin{center}
\begin{tabular}{c|cccccccc}
\hline \hline
   & $M_{H_u}$   &  $M_{H_\phi}$  & $M_{H_1}$  & $M_{H_2}$ & $M_{A^0}$  & $M_{A^{0\, \prime}}$ & $M_{H_1^\pm}$  & $M_{H_2^\pm}$ \\ \hline
 A  & $125\,{\rm GeV}$   &  $13.2\,{\rm TeV}$  & $14.1\,{\rm TeV}$  & $10.7\,{\rm TeV}$  & $13.4\,{\rm TeV}$  & $13.2\,{\rm TeV}$ & $13.2\,{\rm TeV}$  & $7.1 \,{\rm TeV}$  \\  
 B    &  $125\,{\rm GeV}$  &  $29.5\,{\rm TeV}$ &  $50.1\,{\rm TeV}$   &   $33.3\,{\rm TeV}$  &   $2.9\,{\rm TeV}$   &   $2.9\,{\rm TeV}$  &   $2.9\,{\rm TeV}$  &   $15.8\,{\rm TeV}$  \\  \hline
   & $\lambda$   &  $\lambda_1$  & $\lambda_2$  & $\lambda_3$  & $\lambda_4$  & $\lambda_5$ & $\tan\tilde \beta$  & $$  \\  \hline
 A &  $0.51$  & $2.0$ & $1.0$  & $1.0$  & $0.2$ & $0.1$ & $1.7$ & $$ \\ 
  B    & $0.13$   &  $1.0$ & $1.0$   & $1.0$  & $0.1$  & $0.1$   &  $10.0$  &   \\   \hline
    & $\kappa_1$  & $\kappa_2$  & $\kappa_3$  & $\kappa_4$  & $v_u$ & $v_\phi$  & $\nu$  & $V_{331}$  \\  \hline
A &  $0.8$  & $1.0$ & $1.0$  & $1.0$  & $246\,{\rm GeV}$ & $1.0\,{\rm GeV}$ & $-100\,{\rm GeV}$ & $10\,{\rm TeV}$ \\
 B     & $0.2$  & $0.2$  &  $0.2$  & $0.2$  &  $246\,{\rm GeV}$ & $2.0\,{\rm GeV}$   &  $-200\,{\rm GeV}$  & $50\,{\rm TeV}$  \\   \hline
\hline
\end{tabular}
\end{center}
\caption{
The Higgs spectrum and the parameters in the Higgs potential.}
\label{tab:331_Higgs}
\end{table}%

In the end of this section, we briefly summarize the Higgs spectrum in the current context.
The symmetry breaking of $\Gc_{331}\to \Gc_{\rm SM}$ and the sequential EWSB require eight NGBs, while the Higgs sector contains three ${\rm SU}(3)_W$ anti-fundamentals of $\Phi_{\repb{3}\,,\rho}$ and $\Phi_{\repb{3}}^\prime$.
Therefore, we have ten real scalars in all.
Through the above analysis, we find the $331$ Higgs spectrum is consist of: two charged Higgs bosons of $H_{1\,,2}^\pm$ from Eq.~\eqref{eq:331_ChargedHmass_approx}, two CP-odd Higgs bosons of $(A^0\,,  A^{0\,\prime})$ from Eq.~\eqref{eq:331_CPoddmass}, and four CP-even Higgs bosons of $(H_u\,, H_\phi \,, H_1\,, H_2)$ from Eqs.~\eqref{eq:331_CPevenHiggs} and \eqref{eq:331_CPeven_mix}. 
The explicit expressions for Higgs mass matrix in Eqs.~\eqref{eq:331_Charged}, \eqref{eq:331_CPodd}, and \eqref{eq:331_CPevenHiggs} will be given in App.~\ref{sec:331_Hspec}.
At the EW scale, our Higgs spectrum only contains one CP-even Higgs boson of $H_u$, while all other Higgs boson masses are decoupled.
Therefore, our effective theory at the EW scale is distinct from the 2HDM, where a total of four Higgs bosons with masses of the EW scale are generally expected.
Here, we list two benchmark models for the Higgs spectrum in Tab.~\ref{tab:331_Higgs} to demonstrate our results explicitly.
\bigskip

\section{Bottom quark and tau lepton masses in the ${\rm SU}(6)$}
\label{sec:btau}

\subsection{The Yukawa couplings}

By taking the Higgs VEVs in Eqs.~\eqref{eq:331_VEV01} and \eqref{eq:331_VEV02}, we have the following mass matrices for the down-type $(b\,,B)$ quarks and charged $(\tau\,,E)$ leptons
\beqs\label{eqs:SU6Yukawa}
\beqn
&&  (Y_\Dc )_{\rho\sigma } \Big[  ( \mathbf{3}\,, \mathbf{ 3}\,, 0 )_{ \mathbf{F}} \otimes  ( \mathbf{1}\,, \mathbf{ \bar 3}\,,  -\frac{1}{3} )_{ \mathbf{H}}^\rho   \otimes ( \mathbf{\bar 3}\,, \mathbf{ 1}\,, +\frac{1}{3} )_{ \mathbf{F}}^{ \sigma}  \non
&\oplus& (  \mathbf{ 1}\,, \mathbf{\bar 3}\,, +\frac{2}{3} )_{ \mathbf{F}} \otimes  ( \mathbf{1}\,, \mathbf{ \bar 3}\,,  -\frac{1}{3} )_{ \mathbf{H}}^\rho  \otimes (  \mathbf{ 1}\,, \mathbf{\bar 3}\,, -\frac{1}{3} )_{ \mathbf{F}}^{ \sigma }    \Big] + H.c.   \non
&\Rightarrow & (\bar b_L \,, \bar B_L ) \cdot \Mc_\Bc \cdot \left( \ba{c} b_R \\ B_R \\   \ea  \right) + (\bar \tau_L \,, \bar E_L ) \cdot \Mc_\Ec  \cdot \left( \ba{c} \tau_R \\ E_R \\   \ea  \right)  + H.c.  \,, \\
\Mc_\Bc &=& \frac{1}{ \sqrt{2} } \left( \ba{cc}  
(Y_\Dc)_{11} u_1 + ( Y_\Dc )_{12} u_2  & (Y_\Dc)_{21} u_1 + ( Y_\Dc )_{22} u_2  \\
  (Y_\Dc)_{11} V_1 + ( Y_\Dc )_{12} V_2 &  (Y_\Dc)_{21} V_1 + ( Y_\Dc )_{22} V_2 \\   \ea  \right) \sim \left( \ba{cc}  
\epsilon^2  & \epsilon^2  \\
  1 &  1 \\   \ea  \right) V_{331} \,,\label{eq:SU6Yukawa_bottom}\\
\Mc_\Ec &=& \frac{1}{ \sqrt{2} } \left( \ba{cc}  
(Y_\Dc)_{11} u_1 + ( Y_\Dc )_{12} u_2  & -  (Y_\Dc)_{11} V_1 - ( Y_\Dc )_{12} V_2 \\
(Y_\Dc)_{21} u_1 + ( Y_\Dc )_{22} u_2  &  -(Y_\Dc)_{21} V_1 - ( Y_\Dc )_{22} V_2 \\   \ea  \right) \sim \left( \ba{cc}  
\epsilon^2  & -1  \\
  \epsilon^2 &  - 1 \\   \ea  \right) V_{331} \,.\label{eq:SU6Yukawa_tau}
\eeqn
\eeqs
Given the seesaw-like mass matrices according to the mass hierarchy given in Eq.~\eqref{eq:BM_input}, a suppressed bottom quark and tau lepton masses of $\sim\Oc(1)\,{\rm GeV}$ can be realized with the natural Yukawa couplings of $(Y_{\Dc})_{ij} \sim \Oc(1)$.
\bigskip

\subsection{The bottom quark mass}

Specifically, we first illustrate the bottom quark mass, and the tau lepton mass can be obtained straightforwardly.
In general, the mass matrix in Eq.~\eqref{eq:SU6Yukawa_bottom} can be diagonalized through the bi-unitary transformation as
\beqn\label{eq:SU6Yukawa_mixing}
&& U_L^\Bc \Mc_\Bc (U_R^\Bc)^\dag =  \left( \ba{cc}  
m_b  &  0 \\
0  &  m_B \\    \ea  \right) \,,\non
&& U_{L/R}^\Bc =   \left( \ba{cc}  
c_{L/R}  &  -s_{L/R} \\
s_{L/R}  &  c_{L/R}  \\    \ea  \right)   \,, \quad \left( \ba{c}  
b_{L/R}^\prime \\  B_{L/R}^\prime  \ea  \right) = U_{L/R}^\Bc \cdot  \left( \ba{c}  
b_{L/R} \\  B_{L/R}  \ea  \right) \,,
\eeqn
with $(b^\prime\,, B^\prime)$ being the mass eigenstates.
We find that the corresponding Yukawa couplings are expressed in terms of masses and mixing angles as follows
\beqs\label{eqs:SU6Yukawa_coeff}
\beqn
(Y_\Dc )_{11} &=& \sqrt{2} \Big[  ( c_L c_R m_b + s_L  s_R m_B  ) \frac{s_{\tilde \beta} }{ v_\phi }  + ( - s_L c_R m_b + c_L  s_R m_B ) \frac{ c_{\tilde \beta}  }{ V_{331} }  \Big]  \,,\\
(Y_\Dc )_{12} &=& \sqrt{2} \Big[  ( c_L c_R m_b + s_L  s_R m_B ) \frac{ -c_{\tilde \beta} }{ v_\phi }  + ( - s_L c_R m_b + c_L  s_R m_B ) \frac{ s_{\tilde \beta}  }{ V_{331} }   \Big]  \,,\\
(Y_\Dc )_{21} &=& \sqrt{2} \Big[  ( - c_L s_R m_b + s_L  c_R m_B ) \frac{s_{\tilde \beta} }{ v_\phi }  + ( s_L s_R m_b + c_L  c_R m_B  ) \frac{ c_{\tilde \beta}  }{ V_{331} }    \Big]   \,,\\
(Y_\Dc )_{22} &=& \sqrt{2} \Big[ ( - c_L s_R m_b + s_L  c_R m_B ) \frac{ -c_{\tilde \beta} }{ v_\phi }  + ( s_L s_R m_b + c_L  c_R m_B ) \frac{ s_{\tilde \beta}}{ V_{331} }  \Big]   \,.
\eeqn
\eeqs
Under the reasonable limit of $\varphi_{L/R}\to 0$ and $t_{\tilde \beta}\sim 1$, we find the natural Yukawa couplings of $(Y_{\Dc})_{11}\sim (Y_{\Dc})_{12} \sim m_b/v_\phi \sim \Oc(1)$ and $(Y_{\Dc})_{21}\sim (Y_{\Dc})_{22} \sim m_B/V_{331} \sim \Oc(1)$.
\bigskip

By performing the orthogonal transformation in Eq.~\eqref{eq:331CPeven_transf01}, we find the following bottom quark Yukawa couplings with the CP-even Higgs bosons
\beqn
-\Lc_Y^{\Qc\,, 0^+}&\supset& - m_b  \overline{b_L^\prime }  b_R^\prime  \Big[   ( c_L^2 \frac{ c_\theta }{ v_\phi } + s_L^2 \frac{ s_\theta }{ V_{331} }  ) \phi^0  + s_L c_L ( \frac{c_\theta }{ v_\phi } + \frac{ s_\theta }{V_{331} } ) h_1^\prime + ( c_L^2 \frac{ s_\theta }{ v_\phi } - s_L^2 \frac{ c_\theta }{ V_{331} }   )  h_2^\prime \Big]  + H.c. \non
&\approx&  - m_b  \overline{b_L^\prime }  b_R^\prime  \Big( \frac{ c_L^2 }{ v_\phi } \phi^0  + \frac{ s_L c_L }{ v_\phi } h_1^\prime + \frac{ c_L^2 - s_L^2 }{V_{331} } h_2^\prime  \Big) + H.c.\,,
\eeqn
from Eq.~\eqref{eq:bbYuk_CPeven}, and with the approximation given under the decoupling limit of $v_\phi / V_{331} \to 0$.
By using the  orthogonal transformation to CP-even Higgs fields in Eq.~\eqref{eq:331_CPeven_mix}, the bottom quark Yukawa coupling with the SM-like Higgs boson of $H_u$ reads
\beqn\label{eq:SMHbb_Yuk}
-\Lc_Y[H_u]&\supset& - m_b  \overline{b_L^\prime }  b_R^\prime  \Big[ ( c_L^2 \frac{ c_\theta }{ v_\phi } + s_L^2 \frac{ s_\theta }{ V_{331} }  ) (V_{0^+}^{(1)} )_{12}  + s_L c_L  ( \frac{c_\theta }{ v_\phi } + \frac{s_\theta }{ V_{331} }  ) (V_{0^+}^{(1)} )_{13} \non
&+& ( c_L^2 \frac{ s_\theta }{ v_\phi } - s_L^2 \frac{ c_\theta }{ V_{331} }   ) (V_{0^+}^{(1)} )_{14} \Big] H_u + H.c. \non
&\approx&- m_b  \overline{b_L^\prime }  b_R^\prime \Big[ \frac{ c_L^2 }{ v_\phi } (V_{0^+}^{(1)} )_{12}  + \frac{ s_L c_L }{ v_\phi } (V_{0^+}^{(1)} )_{13} + \frac{ c_L^2 - s_L^2 }{V_{331} }  (V_{0^+}^{(1)} )_{14} \Big]  H_u + H.c. \non 
&\approx&- m_b  \overline{b_L^\prime }  b_R^\prime \Big[ \frac{ c_L^2 }{ v_\phi } (V_{0^+}^{(1)} )_{12}  + \frac{ s_L c_L }{ v_\phi } (V_{0^+}^{(1)} )_{13}  \Big]  H_u + H.c.  \,,
\eeqn
with the mixing matrices in Eqs.~\eqref{eq:331_CPeven_mix}, \eqref{eq:331_CPeven_mix_expr}, \eqref{eq:331_CPeven_mix1} for the CP-even Higgs bosons. 
Likewise, we find the Yukawa couplings of the SM-like Higgs boson of $H_u$ with the heavy $B^\prime$ quark as
\beqn\label{eq:SMHBB_Yuk}
-\Lc_Y[H_u]&\approx& - m_B \overline{B_L^\prime }  B_R^\prime \Big[ s_L^2 c_\theta \frac{s_\beta^2 + s_\beta c_\beta }{ v_\phi }  (V_{0^+}^{(1)} )_{12} - s_L c_L \frac{c_\theta }{ v_\phi }  (V_{0^+}^{(1)} )_{13}  \Big] H_u + H.c.\,.
\eeqn
\bigskip

One can expect two constraints from the SM sector, namely, 

\begin{itemize}

\item[$\XG$] the EW charged currents mediated by $W^\pm$,

\item[$\YG$] all SM-like Higgs boson couplings, including $H_u \bar f f$, $H_u VV$ ($V=W^\pm \,,Z$), $H_u gg $, and $H_u \gamma\gamma$.

\end{itemize}
From the EW charged currents given in terms of the gauge eigenstates in Eq.~\eqref{eq:331_EWCC}, it is straightforward to find that $V_{tb} = c_L$.
It is thus natural to take the limit of $c_L\to 1$, according to the measurement of the CKM mixing angle of $|V_{tb}| = 1.013\pm 0.030$~\cite{ParticleDataGroup:2020ssz}.
Under this limit, the SM-like Higgs boson couplings to the heavy $B^\prime$ quark vanishes in Eq.~\eqref{eq:SMHBB_Yuk} as $\varphi_L\to 0$.
Thus, the potential heavy $B^\prime$ quark contributions to the effective $H_u gg$ and $H_u \gamma\gamma$ couplings are vanishing in this limit.
Let us return to the bottom quark Yukawa coupling in Eq.~\eqref{eq:SMHbb_Yuk} when $\varphi_L \to 0$, it is further simplified to
\beqn\label{eq:SMHbb_Yukapprox}
-\Lc_Y[H_u]  &\approx&- \frac{ m_b }{ v_\phi } (V_{0^+}^{(1)} )_{12}\,  \overline{b_L^\prime }  b_R^\prime   H_u + H.c.  \,.
\eeqn
%
%
%
%
%
%
By requiring that the tree-level $H_u \bar b^\prime b^\prime$ Yukawa coupling in Eq.~\eqref{eq:SMHbb_Yuk} is the same as the SM prediction~\cite{ATLAS:2018kot,CMS:2018nsn}, we find the relation of
\beqn\label{eq:SMHbb_Yuk_relation2}
&&  \frac{v_{\rm EW} }{ v_\phi} (V_{0^+}^{(1)} )_{12}  \approx 1 \Rightarrow \frac{ v_{\rm EW}^2 }{ v_\phi V_{331} } \approx 1 \,,
\eeqn
with the mixing angle of $(V_{0^+}^{(1)} )_{12} \sim \epsilon = \frac{v_{\rm EW}}{V_{331}}$ in Eq.~\eqref{eq:331_CPeven_mix1}.
For simplicity, the sub-leading correction term suppressed by $1/V_{331}$ in Eq.~\eqref{eq:SMHbb_Yuk} is neglected.
Apparently, this relation leads to the natural new physics scale for the $331$ symmetry of 
\beqn\label{eq:V331}
&& \boxed{ V_{331} \sim  \Oc(10)\,\TeV } \,,
\eeqn
with the reasonable choice of $v_\phi \sim \Oc(1)\,{\rm GeV}$ for the bottom quark Yukawa coupling.
This confirms our previous assumption of the benchmark parameter input in Eq.~\eqref{eq:BM_input}.
We have also checked that a new physics scale of $V_{331}$ in Eq.~\eqref{eq:V331} is even consistent with the most stringent limit to the rare flavor-changing lepton decay process of ${\rm Br}(\mu \to e \gamma)$~\cite{MEG:2016leq} when generalizing to the three-generational case~\cite{Li:2019qxy}.
\bigskip

\subsection{The tau lepton mass}

The tau lepton mass and Yukawa couplings follow closely from the bottom quark case, and we present the discussion here for completeness.
The general $\Ec=(\tau\,, E)$ mass matrix in Eq.~\eqref{eq:SU6Yukawa_tau} is related to the $\Bc=(b\,,B)$ mass matrix in Eq.~\eqref{eq:SU6Yukawa_bottom} by
\beqn
\Mc_\Ec&=& \Mc_\Bc^T \cdot \sigma_3 \,.
\eeqn
It is straightforward to find that the bi-unitary transformation for the $\Ec=(\tau\,, E)$ is simply related to those for the $\Bc=(b\,,B)$ as below
\beqn\label{eq:SU6Yukawa_mixing_tau}
&& U_L^\Ec \Mc_\Ec (U_R^\Ec )^\dag =  \left( \ba{cc}  
m_\tau  &  0 \\
0  &  m_E \\    \ea  \right) \,,\quad  \left( \ba{c}  
\tau_{L/R}^\prime \\  E_{L/R}^\prime  \ea  \right) = U_{L/R}^\Ec \cdot  \left( \ba{c}  
\tau_{L/R} \\  E_{L/R}  \ea  \right) \non
&& U_{L}^\Ec =   \left( \ba{cc}  
c_{R}  &  -s_{R} \\
s_{R}  &  c_{R}  \\    \ea  \right)   \,, \quad 
U_{R}^\Ec =   \left( \ba{cc}  
c_{L}  &  s_{L} \\
-s_{L}  &  c_L  \\ \ea  \right)   \,.
\eeqn
Immediately, this leads a result of $s_R=0$ from leptonic sector of the EW charged currents in Eq.~\eqref{eq:331_EWCC}.
Analogous to Eqs.~\eqref{eqs:SU6Yukawa_coeff}, the Yukawa couplings can also be expressed as
\beqs\label{eqs:SU6Yukawa_coeff_tau}
\beqn
(Y_\Dc )_{11} &=& \sqrt{2} \Big[  ( c_L c_R m_\tau - s_L  s_R m_E  ) \frac{s_{\tilde \beta} }{ v_\phi }  + ( s_L c_R m_\tau + c_L  s_R m_E ) \frac{ - c_{\tilde \beta}  }{ V_{331} }  \Big]  \,,\\
(Y_\Dc )_{12} &=& \sqrt{2} \Big[  ( c_L c_R m_\tau - s_L  s_R m_E ) \frac{ -c_{\tilde \beta} }{ v_\phi }  + (  s_L c_R m_\tau + c_L  s_R m_E ) \frac{ -s_{\tilde \beta}  }{ V_{331} }   \Big]  \,,\\
(Y_\Dc )_{21} &=& \sqrt{2} \Big[  ( - c_L s_R m_\tau - s_L  c_R m_E ) \frac{s_{\tilde \beta} }{ v_\phi }  + ( s_L s_R m_\tau - c_L  c_R m_E  ) \frac{ c_{\tilde \beta}  }{ V_{331} }    \Big]   \,,\\
(Y_\Dc )_{22} &=& \sqrt{2} \Big[ ( c_L s_R m_\tau + s_L  c_R m_E ) \frac{c_{\tilde \beta} }{ v_\phi }  + ( s_L s_R m_\tau - c_L  c_R m_E ) \frac{ s_{\tilde \beta}}{ V_{331} }  \Big]   \,.
\eeqn
\eeqs
Obviously, Eqs.~\eqref{eqs:SU6Yukawa_coeff} and Eqs.~\eqref{eqs:SU6Yukawa_coeff_tau} lead to the degenerate fermion mass predictions of $m_b = m_\tau$ and $m_B = -m_E$~\footnote{The relative negative sign in $m_B = -m_E$ can always be rotated away by redefining the right-handed component of $E$.}.
Thus, the $b$-$\tau$ mass unification issue cannot be addressed at the tree level.
Their mass splitting can be attributed to the renormalization group running. 
This was first discussed in the context of the Georgi-Glashow ${\rm SU}(5)$ model~\cite{Buras:1977yy}.
However, results therein cannot be naively applied to the $(b\,,\tau)$ mass ratio in the non-minimal GUTs.
To fully evaluate their mass splitting, we expect two prerequisites of: (i) evaluation of the intermediate symmetry breaking scales from an appropriate GUT group, and (ii) the identification of the SM fermion representations with the extended color and weak symmetries.
Both are distinctive features of the non-minimal GUTs, and we defer to analyze the details in the future work.
By performing the orthogonal transformation in Eq.~\eqref{eq:331CPeven_transf01}, we find the tau lepton Yukawa coupling with the SM-like Higgs boson the same as that for the bottom quark case in Eq.~\eqref{eq:SMHbb_Yuk}, with $m_b \to m_\tau$.
Therefore, the scale of $V_{331}$ in Eq.~\eqref{eq:V331} can be similarly determined from the tau lepton, given the current LHC measurements of the $H_u \tau \tau$ coupling~\cite{CMS:2017zyp,ATLAS:2018ynr}.
\bigskip

\subsection{The possible radiative mechanism}

We comment on the possible radiative fermion mass generation in the current scenario, which was proposed and considered to produce fermion mass hierarchies in various context~\cite{Georgi:1972mc,Georgi:1972hy,Barr:1976bk,Barr:1978rv,Barr:1979xt,Ibanez:1981nw,Ma:2006km,Dobrescu:2008sz,Weinberg:2020zba}.
In such a paradigm, the general assumption is that some light fermion masses can be radiatively generated with vanishing tree-level masses.
Specifically, we should check that whether the bottom quark and tau lepton masses can be generated with $m_b=0$ in Eq.~\eqref{eq:SU6Yukawa_mixing} and $m_\tau=0$ in Eq.~\eqref{eq:SU6Yukawa_mixing_tau}.
Let us consider the $\Bc=(b\,, B)$ case without loss of generality, the Yukawa couplings are reduced to the following expressions
\beqs
\beqn
(Y_\Dc )_{11} &=& \frac{m_B}{ v_\phi } s_R ( s_L s_{\tilde \beta} + t_\theta c_L c_{\tilde \beta} )  \,,\\
(Y_\Dc )_{12} &=& \frac{m_B}{ v_\phi } s_R ( t_\theta c_L s_{\tilde \beta} -  s_L c_{\tilde \beta} ) = (Y_\Dc )_{22} t_R\,,\\
(Y_\Dc )_{21} &=& \frac{m_B}{ v_\phi } c_R ( s_L s_{\tilde \beta} + t_\theta c_L c_{\tilde \beta} ) =  (Y_\Dc )_{11} / t_R \,,\\
(Y_\Dc )_{22} &=& \frac{m_B}{ v_\phi } c_R ( t_\theta c_L s_{\tilde \beta} -  s_L c_{\tilde \beta} )  \,,
\eeqn
\eeqs
under the vanishing tree-level mass of $m_b=0$.
The bottom quark and its heavy partner $B$ can be mediated through the flavor-changing neutral vector bosons of $(N_\mu \,, \bar N_\mu )$ as in Eq.~\eqref{eq:331_FCNC}, while this only happens for the left-handed components.
Thus, the neutral vector bosons of $(N_\mu \,, \bar N_\mu)$ cannot lead to a radiative mass terms as was suggested in Refs.~\cite{Barr:1976bk,Barr:1978rv}.
The remaining possibility may be due to the mediation from the Higgs sector, as in Ref.~\cite{Ma:2006km}.
By taking the $m_b=0$ in Eqs.~\eqref{eq:bbYuk_CPodd} and \eqref{eq:bbYuk_CPeven}, the neutral Higgs bosons can only mediate the left-handed $b^\prime$ and right-handed $B^\prime$.
Thus, it is impossible to generate a radiative mass term of $ m_b^{\rm rad} \overline{b_L^\prime } b_R^\prime + H.c.$ with a vanishing tree-level $m_b=0$.
The same argument also applies to the $\Ec=(\tau\,, E)$ case with the $m_\tau=0$ limit.
\bigskip

\section{Conclusions}
\label{sec:conclusion}
%
%

In this work, we study the bottom quark and tau lepton mass generations in the framework of one-generational ${\rm SU}(6)$ GUT.
The symmetry breaking stage of $\Gc_{331}\to \Gc_{\rm SM}$ is found to be general for more realistic non-minimal GUTs with $n_g=3$.
A different assignment to the Higgs VEVs from the previous studies is considered so that the bottom quark and tau lepton can obtain tree-level masses with $\sim\Oc(1)$ Yukawa couplings.
We consider this fermion-Higgs mismatching pattern to be general, such as in more realistic unified model with the ${\rm SU}(8)$ symmetry.
Thus, we prevent the pattern leading to multiple EW Higgs doublets, which is very problematic with the ongoing LHC searches.
An automatically generated small Higgs VEVs of $\sim\Oc(1)\,\GeV$ is found to be possible as long as a gauge-invariant $\nu$-term in the Higgs potential can be of $\sim \Oc(100)\,\GeV$.
Notice that this $\nu$-term is also invariant under the emergent global symmetry of Eq.~\eqref{eq:SU6_flavor}, which emerges automatically from the anomaly-free condition.
By requiring the Yukawa coupling of the SM-like Higgs boson to SM fermions of $y_f^{\rm SM}=\frac{\sqrt{2} m_f}{v_{\rm EW} }$, we find the $331$ symmetry-breaking scale of $V_{331}\sim 10\,\TeV$ in the current context.
This was not mentioned in the previous context.
With the distinct VEV assignments in Eqs.~\eqref{eq:331_VEV01} and \eqref{eq:331_VEV02}, we find a Higgs sector consisting of one single CP-even Higgs boson at the EW scale.
All other Higgs bosons have masses of $\sim{\cal O}(V_{331})$, as we have described in Sec.~\ref{sec:SU6_Higgs}.
Therefore, the effective theory at the EW scale contains only one SM-like CP-even Higgs boson, and is not described by a 2HDM.
\bigskip

Historically, it was proposed that three-generational SM fermions may be embedded non-trivially in a non-minimal GUT~\cite{Georgi:1979md}.
Through our recent analysis~\cite{Chen:2021ovs,Chen:2022xge}, we find that the ${\rm SU}(8)$ GUT can be the minimal model that have three-generational SM fermions transform differently under the extended gauge symmetries beyond the EW scale.
Through the current discussion, we wish to mention the relations between the ${\rm SU}(6)$ toy model and the realistic ${\rm SU}(8)$ model.
First, the ${\rm SU}(6)$ subgroup of the ${\cal G}_{331}$ can be generic in the context of the ${\rm SU}(8)$ GUT, as was shown in Eq.~\eqref{eq:SU8_Pattern}.
Therefore, the results such as the ${\cal G}_{331}$ gauge sector and part of the Higgs sector in the current discussion can become useful in the context of the ${\rm SU}(8)$ model.
Second, the symmetry breaking pattern can be generalized, where the seemingly unnatural  $\nu$-term in Eq.~\eqref{eq:2HTM_potential03} that generates the EWSB VEVs for the $(b\,,\tau)$ masses are natural due to the gravitational effect in the ${\rm SU}(8)$ model.
This means a potential relation between the gravitational effect and the flavor sector, which was never mentioned in any previous GUT literature according to our knowledge.
Since the one-generational ${\rm SU}(6)$ GUT is a toy model, there are several issues beyond the scope of the current discussions.
They include: (i) the $b$-$\tau$ mass unification, (ii) the three-generational SM fermion mixings.
Furthermore, the SM fermions in the non-minimal GUTs are usually accompanied with heavy partners from the ${\rm SU}(N)$ anti-fundamentals.
They can be mediated through the heavy charged and/or neutral vector bosons as well as heavy Higgs bosons during the intermediate symmetry breaking stages of the non-minimal GUT symmetry.
It is therefore necessary to carry out detailed analysis of their experimental implications in some rare flavor-changing processes.
All these issues will be studied elsewhere when extending to more realistic non-minimal GUTs such as the ${\rm SU}(8)$, where three-generational SM fermions are embedded non-trivially.
\bigskip

\section*{Acknowledgements}

We would like to thank Wenbin Yan, Chang-Yuan Yao, and Ye-Ling Zhou for enlightening discussions. 
N.C. would like to thank Tibet University, Shanghai Institute of Optics and Fine Mechanics (CAS), and Peking University for hospitality when preparing this work.
N.C. is partially supported by the National Natural Science Foundation of China (under Grants No. 12035008 and No. 12275140), and Nankai University.
Y.N.M. is partially supported by the National Natural Science Foundation of China (under Grant No. 12205227), the Fundamental Research Funds for the Central Universities (WUT: 2022IVA052), and Wuhan University of Technology. 
\bigskip

\appendix

\section{The gauge symmetry breaking in the $331$ model}
\label{sec:331_gauge}

In this section, we summarize the necessary results of the gauge symmetry breaking of ${\cal G}_{331} \to {\cal G}_{\rm SM}$ for the current discussions as well as for the future studies.
\bigskip

\subsection{The $331$ gauge bosons}
\label{sec:331_GB}

 The kinematic terms for the ${\rm SU}(3)_W$ Higgs fields are
 \beqn
 \Lc&=& \sum_\rho |D_\mu \Phi_{ \repb{3} \,,\rho} |^2 +  |D_\mu \Phi_{ \repb{3} }^\prime |^2 \,.
 \eeqn
Generically, the covariant derivative for the ${\rm SU}(3)_W$ fundamental representation is defined according to the convention in Refs.~\cite{Ferreira:2011hm,Cao:2016uur}~\footnote{In some $331$ literatures, e.g. Ref.~\cite{Buras:2012dp}, the ${\rm U}(1)$ charge is defined with a $3\times 3$ unity matrix of $\frac{1}{ \sqrt{6}} \mathbb{I}_3$.}
\beqn\label{eq:31cov_fund}
&& D_\mu \Phi_{ \rep{3} } \equiv  ( \partial_\mu - i g_{3L} A_\mu^a \frac{\lambda^a}{2} -  i g_X X  \mathbb{I}_3 X_\mu )  \Phi_{ \rep{3} } \,,
\eeqn
with $\lambda^a\,(a=1\,,...\,,8)$ being the ${\rm SU}(3)$ Gell-Mann matrices.
For the ${\rm SU}(3)_W$ anti-fundamental representation, the covariant is defined as
\beqn\label{eq:31cov_antifund}
&& D_\mu \Phi_{ \repb{3} } \equiv  ( \partial_\mu + i g_{3L} A_\mu^a \frac{( \lambda^a)^* }{2} -  i g_X X  \mathbb{I}_3 X_\mu )  \Phi_{ \repb{3} } \non
&=& ( \partial_\mu + i g_{3L} A_\mu^a \frac{( \lambda^a)^T }{2} -  i g_X X  \mathbb{I}_3 X_\mu )  \Phi_{ \repb{3} }  \,,
\eeqn
with the hermiticity of $(\lambda^a)^\dag = \lambda^a$.
Note that the definitions in Eqs.~\eqref{eq:31cov_fund} and \eqref{eq:31cov_antifund} are applicable to the ${\rm SU}(3)_W$ fermions.
\bigskip

Explicitly, we express the gauge fields in terms of a $3\times 3$ matrix as follows
\beqn\label{eq:331_connection}
&&  g_{ 3 L } A_\mu^a \frac{\lambda^a}{2} + g_X X  \mathbb{I}_3 X_\mu \non 
&=& \small{ \frac{ 1 }{2} \left( \ba{ccc}  
  g_{3 L}  ( A_\mu^3 + \frac{1}{\sqrt{3} } A_\mu^8 ) + 2 g_X X X_\mu &  g_{3L} (A_\mu^1 - i A_\mu^2  ) &  g_{3L} (A_\mu^4 - i A_\mu^5  )  \\ 
  g_{3L} (A_\mu^1 +  i A_\mu^2  ) &  g_{3 L}  ( - A_\mu^3 + \frac{1}{\sqrt{3} } A_\mu^8 ) +  2 g_X X X_\mu  & g_{3L} (A_\mu^6 - i A_\mu^7  )   \\ 
 g_{3L}  (A_\mu^4 + i A_\mu^5  ) & g_{3L} (A_\mu^6 + i A_\mu^7  )  &   - \frac{ 2 g_{3 L}}{\sqrt{3} }  A_\mu^8  +  2 g_X X X_\mu   \\   \ea  \right) } \non
  && \,.
\eeqn
One can identify the charged gauge bosons of $W_\mu^\pm \equiv \frac{1}{ \sqrt{2} } (A_\mu^1 \mp i A_\mu^2 ) $, $C_\mu^\pm \equiv \frac{1}{ \sqrt{2} } (A_\mu^4 \mp i A_\mu^5 ) $, and the neutral gauge bosons of $N_\mu \equiv \frac{1}{ \sqrt{2} } (A_\mu^6 - i A_\mu^7 ) $, and $\bar N_\mu \equiv \frac{1}{ \sqrt{2} } (A_\mu^6 + i A_\mu^7 ) $.
The electric charges of gauge bosons can be obtained by the relation of $[\hat Q\,, A_\mu^a \lambda^a ] = Q_A^{IJ}  (A_\mu^a \lambda^a)_{IJ}$, with $X=0$ (since the ${\rm SU}(3)_W$ gauge bosons do not take the ${\rm U}(1)_X$ charges) in the electric charge operator given in Eq.~\eqref{eq:Qcharge_331}.
\bigskip

The charged and neutral $331$-gauge boson mass squares at the tree level read
\beqn
&& m_{C_\mu^\pm}^2 = m_{N_\mu\,, \bar N_\mu }^2 = \frac{1}{4} g_{3L}^2 V_{331}^2  \,,
\eeqn
with the VEV assignment in Eq.~\eqref{eq:331_VEV01} for simplity.
The other neutral gauge boson is due to the linear combination of $(A^8_\mu \,, X_\mu)$, whose mass matrix is
\beqn
&& \hf \cdot \frac{ (V_{331} )^2 }{9} ( A^{8\, \mu }\,, X^\mu  )  \left( \ba{cc}  
   3 g_{3L}^2 & - \sqrt{3} g_{3L} g_X  \\ 
 - \sqrt{3} g_{3L} g_X   &   g_X^2   \\   \ea \right) \cdot \left( \ba{c}  A^8_\mu  \\  X_\mu  \\    \ea \right) \,, \non
 &\Rightarrow& m_{Z_\mu^\prime}^2 = \frac{1}{9}( g_X^2  + 3 g_{3L}^2 ) (V_{331} )^2  \,.
\eeqn
It is straightforward to define a mixing angle $\theta_X$ for the $331$ symmetry breaking as
\beqn
 \left( \ba{c}  Z^\prime_\mu  \\ B_\mu  \\    \ea \right) &=& \left( \ba{cc}  
   c_{X}  &  - s_{X} \\ 
 s_{X}  & c_{X}   \\   \ea \right) \cdot  \left( \ba{c}  A^8_\mu  \\ X_\mu \\     \ea \right)\,,
\eeqn
with
\beqn
t_X&\equiv& \tan\theta_{X} = \frac{g_X }{ \sqrt{3} g_{3L} }\,.
\eeqn
Thus, $Z^\prime_\mu$ and $B_\mu$ can be expressed in terms of $A_\mu^8$ and $X_\mu$ as
\beqn
&& Z^\prime_\mu = \frac{ \sqrt{3} g_{3L} A_\mu^8 -  g_X X_\mu }{ \sqrt{  g_X^2 + 3 g_{3L}^2 }} \,, \quad  B_\mu = \frac{  g_{X} A_\mu^8 +  \sqrt{3} g_{3L} X_\mu }{ \sqrt{  g_X^2 + 3 g_{3L}^2 }  } \,.
\eeqn
The ${\rm U}(1)_Y$ coupling of $\alpha_Y$ is related to the ${\rm SU}(3)_W \otimes {\rm U}(1)_X$ couplings of $(\alpha_{3L}\,, \alpha_X )$ as
\beqn
&& \alpha_Y^{-1} = \frac{1}{3} \alpha_{3L }^{-1} +  \alpha_X^{-1}\,, \quad \frac{1}{3} \alpha_{3L}^{-1} = \alpha_Y^{-1} s_{X}^2 \,,\quad \alpha_X^{-1} = \alpha_Y^{-1} c_{X}^2  \,.
\eeqn
Correspondingly, the diagonal components of the ${\rm SU}(3)_W \otimes {\rm U}(1)_X$ covariant derivative in Eq.~\eqref{eq:331_connection} become
\beqn
 &&  \hf {\rm diag} \Big( g_{3L} A_\mu^3 + g_Y (\frac{1}{3} + 2X ) B_\mu \,, - g_{3L} A_\mu^3 + g_Y (\frac{1}{3} + 2X ) B_\mu   \,,   g_Y (-\frac{2}{3} + 2X ) B_\mu  \Big) \non
&+& \frac{g_Y}{6} {\rm diag} \Big(  -6X t_{X} + \frac{1}{t_X} \,, -6X t_{X} + \frac{1}{t_X}\,,  -6X t_{X} - \frac{2}{t_X}  \Big) Z^\prime_\mu \,.
\eeqn
Clearly, the $A_\mu^3$ and $B_\mu$ terms from first two components recover the covariant derivatives in the EW theory with $X=\frac{1}{3}$.
The off-diagonal components in Eq.~\eqref{eq:331_connection} become
\beqn
&& \frac{g_{3L} }{ \sqrt{2}} \left( \ba{ccc}  
  0 &  W_\mu^+ &  C_\mu^+  \\ 
 W_\mu^- &  0  &  N_\mu  \\ 
C_\mu^-  &  \bar N_\mu &  0  \\   \ea  \right)\,.
\eeqn
Below, we call five massive gauge bosons of
\beqn
&& \Big\{  C_\mu^\pm  \,, N_\mu \,, \bar N_\mu \,, Z_\mu^\prime  \Big\} 
\eeqn
at this stage of symmetry breaking as the $331$ gauge bosons, while the remaining ones of $ \Big\{  W_\mu^\pm  \,, A_\mu^3 \,, B_\mu \Big\}$ are the usual EW gauge bosons.
\bigskip

\subsection{The gauge couplings of fermions}
\label{sec:331_fgauge}

The ${\rm SU}(3)_W \otimes {\rm U}(1)_X$ covariant derivatives for chiral fermions in Tab.~\ref{tab:SU6_331_ferm} are expressed as follows in terms of gauge eigenstates
 \beqs
 \beqn
 && i  D_\mu \Bc_R^\rho \supset     g_X (- \frac{1}{3}  )  X_\mu \Bc_R^\rho \,,\\
 && i  D_\mu \Lc_L^\rho \supset  (  - g_{3L} A_\mu^a \frac{ (\lambda^a)^T }{2} + g_X X_\mu \mathbb{I}_3 ( -\frac{1}{3})   ) \Lc_L^\rho \,,\\
 &&i D_\mu t_R \supset  g_X (+\frac{2}{3} ) X_\mu  t_R \,,\\
 && i  D_\mu \Ec_R \supset  (  + g_{3L} A_\mu^a \frac{ \lambda^a }{2} - g_X X_\mu \mathbb{I}_3 \frac{2}{3}  ) \Ec_R \,,\\
 && i  D_\mu \Qc_L \supset    g_{3L} A_\mu^a \frac{ \lambda^a }{2}  \Qc_L  \,.
 \eeqn
 \eeqs
Analogous to the SM, we should find the charged currents and neutral currents for the ${\rm SU}(3)_W \otimes {\rm U}(1)_X$ gauge bosons. 
 \bigskip

 The flavor-changing ${\rm SU}(3)_W \otimes {\rm U}(1)_X$ charged currents are mediated by $C_\mu^\pm$ as follows
 \beqn\label{eq:331_FCCC}
 \Lc_{ {\rm SU}(3)_W}^{\rm CC}&=& \frac{g_{3L} }{ \sqrt{2} } \Big[ \overline t_L \gamma^\mu B_L + \overline{ N_R}  \gamma^\mu \tau_R -  \overline{\widetilde N_L^1}  \gamma^\mu \tau_L - \overline{\widetilde N_L^2}  \gamma^\mu E_L  \Big] C_\mu^+   + H.c. \,.
 \eeqn
 The ${\rm SU}(3)_W \otimes {\rm U}(1)_X$ neutral currents contain both the flavor-changing components mediated by $(N_\mu \,, \bar N_\mu)$, and the flavor-conserving components mediated by $Z^\prime_\mu$.
 In the chiral basis, they read
 \beqs
 \beqn
  \Lc_{ {\rm SU}(3)_W}^{ {\rm NC\,, \bcancel{\rm F}} }&=&  \frac{g_{3L} }{ \sqrt{2} } \Big[ \overline{\widetilde N_L^1} \gamma^\mu \nu_L  +  \overline{ \widetilde N_L^2 } \gamma^\mu N_L  + \overline{E}_R \gamma^\mu \tau_R + \overline b_L \gamma^\mu B_L \Big] N_\mu \non
  &+&  \frac{g_{3L} }{ \sqrt{2} }  \Big[  \overline{ \nu}_L \gamma^\mu \widetilde N_L^1 +   \overline{ N }_L \gamma^\mu \widetilde N^2_L  + \overline{\tau}_R \gamma^\mu E_R + \overline{ B_L} \gamma^\mu b_L \Big] \bar N_\mu \,\label{eq:331_FCNC} \\
 \Lc_{ {\rm SU}(3)_W}^{\rm NC\,, F}&=& g_Y  \Big[  \overline t_L \gamma^\mu (\frac{1}{6 t_X }    )  t_L + \overline b_L \gamma^\mu (\frac{1}{6 t_X }   )   b_L + \overline B_L \gamma^\mu  ( - \frac{1}{3 t_X }  )  B_L \non
  &+& \overline t_R \gamma^\mu (-\frac{2}{3} t_X ) t_R + \overline b_R \gamma^\mu (\frac{1}{3} t_X  ) b_R + \overline B_R \gamma^\mu (\frac{1}{3} t_X ) B_R \non
  &+& \overline \tau_L \gamma^\mu ( \frac{1 }{3} t_X - \frac{1}{6 t_X}  ) \tau_L + \overline \nu_L \gamma^\mu (\frac{1 }{3} t_X  - \frac{1}{6 t_X } )  \nu_L \non
  &+& \overline E_L \gamma^\mu ( \frac{1 }{3} t_X - \frac{1}{6 t_X} ) E_L + \overline N_L \gamma^\mu ( \frac{1 }{3} t_X - \frac{1}{6 t_X}  ) N_L +\overline N_R \gamma^\mu ( \frac{2}{3} t_X + \frac{1}{6 t_X}  ) N_R \non
  &+& \overline{\widetilde N_L^1} \gamma^\mu ( \frac{1 }{3} t_X + \frac{1}{3 t_X} ) \widetilde N_L^1 +  \overline{\widetilde N^2_L} \gamma^\mu ( \frac{1 }{3} t_X  + \frac{1}{3 t_X} ) \widetilde N^2_L  \non
  &+&\overline E_R \gamma^\mu ( \frac{2}{3} t_X + \frac{1}{6 t_X}  ) E_R +\overline \tau_R \gamma^\mu ( \frac{2}{3} t_X -  \frac{1}{3 t_X}  )  \tau_R \Big] Z_\mu^\prime \,. \label{eq:331_FcCNC}
 \eeqn
 \eeqs
 \bigskip

Apparently, the EW charged currents should reproduce the SM case, which are
\beqn\label{eq:331_EWCC}
  \Lc_{ {\rm SU}(2)_W}^{ {\rm CC} }&=& \frac{g_{3L} }{ \sqrt{2} } \Big[ \overline t_L \gamma^\mu b_L  + \overline \tau_L \gamma^\mu \nu_L \Big] W_\mu^+   + H.c.
\eeqn
\bigskip

\subsection{The mass matrices of the Higgs bosons}
\label{sec:331_Hspec}

The matrix elements for the charged Higgs bosons in Eq.~\eqref{eq:331_Charged} read
\beqs\label{eq:331_Charged_elem} 
\beqn
M_{\phi_1^+ \phi_1^- }^2 &=& \frac{v_u \Big[  \sqrt{2} \nu  \left( s_{\tilde \beta}^2 (v_{\phi}^2 - V_{331}^2)  -v_{\phi}^2 \right)+\kappa_{3} V_{331} v_{\phi} v_u \Big] }{2 V_{331} v_{\phi}}   \,,\\[1mm]
M_{\phi_1^\pm \phi_2^\mp }^2 &=& \frac{  \nu  v_u s_{\tilde \beta } c_{\tilde \beta } (V_{331}^2 - v_{\phi}^2 )  }{ \sqrt{2} V_{331} v_{\phi}}   \,,\\[1mm]
M_{\phi_2^+ \phi_2^- }^2 &=& \frac{ v_u \Big[ \sqrt{2} \nu  \left(c_{\tilde\beta}^2 (v_{\phi}^2 - V_{331}^2) - v_{\phi}^2 \right)+\kappa_{4} V_{331} v_{\phi} v_u \Big] }{2 V_{331} v_{\phi}}   \,,\\[1mm]
M_{\phi_1^\pm \chi_1^\mp }^2 &=& \frac{1}{2} s_{\tilde \beta }  (  \kappa_{3} v_{\phi} v_u-\sqrt{2}  \nu  V_{331}  )  \,,\\[1mm]
M_{\phi_1^\pm \chi_2^\mp }^2 &=& \frac{1}{2} c_{\tilde \beta} (  \kappa_{3} V_{331} v_u-\sqrt{2} \nu  v_{\phi} )   \,,\\[1mm]
M_{\phi_2^\pm \chi_1^\mp }^2 &=& \frac{1}{2} c_{\tilde \beta} ( \sqrt{2} \nu  V_{331} - \kappa_{4} v_{\phi} v_u )  \,,\\[1mm]
M_{\phi_2^\pm \chi_2^\mp }^2 &=& \frac{1}{2} s_{\tilde \beta} (  \kappa_{4} V_{331} v_u-\sqrt{2} \nu  v_{\phi} ) \,,\\[1mm]
M_{\chi_1^+ \chi_1^- }^2 &=& \frac{ v_{\phi}  (-\sqrt{2} \nu  V_{331}+\kappa_{3} v_{\phi} v_u s_{\tilde \beta}^2+\kappa_{4} v_{\phi} v_u c_{\tilde \beta}^2  )  }{2 v_u}   \,,\\[1mm]
M_{\chi_1^\pm  \chi_2^\mp }^2 &=& \frac{1}{4} V_{331} v_{\phi} s_{2 \tilde \beta} (\kappa_{3} - \kappa_{4})   \,,\\[1mm]
M_{\chi_2^+  \chi_2^- }^2 &=& \frac{V_{331} (\kappa_{3} V_{331} v_u c_{\tilde \beta}^2 + \kappa_{4} V_{331} v_u s_{\tilde \beta}^2 - \sqrt{2} \nu v_{\phi} )}{2 v_u}   \,.
\eeqn
\eeqs
The matrix elements for the CP-odd Higgs bosons in Eq.~\eqref{eq:331_CPodd} read
\beqs\label{eq:331_CPodd_elem} 
\beqn
M_{\pi_u \pi_u}^2 &=& -\frac{\nu V_{331} v_{\phi}}{\sqrt{2} v_u}   \,,\\[1mm]
M_{\pi_u \eta_1}^2 &=& \frac{\nu V_{331} s_{\tilde \beta}} {\sqrt{2}}   \,,\\[1mm]
M_{\pi_u \eta_2}^2 &=& -\frac{\nu V_{331} c_{\tilde \beta}}{\sqrt{2}}   \,,\\[1mm]
M_{\pi_u \pi_1}^2 &=& \frac{\nu v_{\phi} c_{\tilde \beta}}{\sqrt{2}}   \,,\\[1mm]
M_{\pi_u \pi_2}^2 &=& \frac{\nu v_{\phi} s_{\tilde \beta}}{\sqrt{2}}   \,,\\[1mm]
M_{\eta_1 \eta_1}^2 &=& \frac{1}{2} v_{\phi}^2 c_{\tilde \beta}^2 (\lambda_{4}-\lambda_{5})-\frac{\nu v_u \Big[   s_{\tilde \beta}^2 (V_{331}^2 - v_{\phi}^2 ) +v_{\phi}^2  \Big] }{\sqrt{2} V_{331} v_{\phi}}   \,,\\[1mm]
M_{\eta_1  \eta_2}^2 &=& \frac{s_{\tilde \beta} c_{\tilde \beta} \Big[  \sqrt{2} \nu ( V_{331}^2 -v_{\phi}^2 ) v_u+V_{331} v_{\phi}^3 (\lambda_{4} - \lambda_{5})  \Big] }{2 V_{331} v_{\phi}}   \,,\\[1mm]
M_{\pi_1 \eta_1}^2 &=& \frac{1}{4} V_{331} v_{\phi} s_{2\tilde \beta} (\lambda_{5} - \lambda_{4})   \,,\\[1mm]
M_{\pi_2 \eta_1}^2 &=& \frac{1}{2} \Big[  V_{331} v_{\phi} c_{\tilde \beta}^2 (\lambda_{4}-\lambda_{5})-\sqrt{2} \nu v_u  \Big]  \,,\\[1mm]
M_{\eta_2 \eta_2}^2 &=& \frac{\nu v_u \Big[  c_{\tilde \beta}^2 (v_{\phi}^2 - V_{331}^2 ) -v_{\phi}^2 \Big]  }{\sqrt{2} V_{331} v_{\phi}}+\frac{1}{2} v_{\phi}^2 s_{\tilde \beta}^2 (\lambda_{4}-\lambda_{5})   \,,\\[1mm]
M_{\pi_1 \eta_2}^2 &=& \frac{1}{2} V_{331} v_{\phi} s_{\tilde \beta}^2 (\lambda_{5}-\lambda_{4})+\frac{\nu v_u}{\sqrt{2}}   \,,\\[1mm]
M_{\pi_2 \eta_2}^2 &=& \frac{1}{4} V_{331} v_{\phi} s_{2\tilde \beta} (\lambda_{4}-\lambda_{5})  \,,\\[1mm]
M_{\pi_1  \pi_1}^2 &=& \frac{1}{2} V_{331}^2 s_{\tilde \beta}^2 (\lambda_{4}-\lambda_{5})+\frac{\nu v_u \Big[ s_{\tilde \beta}^2 (v_{\phi}^2 -V_{331}^2 )  -v_{\phi}^2  \Big]  }{\sqrt{2} V_{331} v_{\phi}}   \,,\\[1mm]
M_{\pi_1 \pi_2}^2 &=& \frac{s_{\tilde \beta} c_{\tilde \beta} \Big[ V_{331}^3 v_{\phi} (\lambda_{5}-\lambda_{4})+\sqrt{2} \nu ( V_{331}^2 - v_{\phi}^2 ) v_u  \Big] }{2 V_{331} v_{\phi}}   \,,\\[1mm]
M_{\pi_2 \pi_2}^2 &=& \frac{1}{2} V_{331}^2 c_{\tilde \beta}^2 (\lambda_{4}-\lambda_{5})+\frac{\nu v_u \Big[ c_{\tilde \beta}^2 (v_{\phi}^2 -V_{331}^2 )  -v_{\phi}^2 \Big] }{\sqrt{2} V_{331} v_{\phi}}   \,.
\eeqn
\eeqs
The matrix elements for the CP-even Higgs bosons in Eq.~\eqref{eq:331_CPevenHiggs} read
\beqs\label{eq:331_CPeven_elem} 
\beqn
M_{h_u  h_u}^2 &=& 2 \lambda v_u^2 - \frac{\nu V_{331} v_{\phi}}{\sqrt{2} v_u}   \,,\\[1mm]
M_{h_u \phi^0 }^2 &=& -\frac{\sqrt{2} \nu V_{331}^2+2 V_{331} v_{\phi} v_u (\kappa_{1}+\kappa_{2})+\sqrt{2} \nu v_{\phi}^2}{2 \sqrt{V_{331}^2+v_{\phi}^2}}   \,,\\[1mm]
M_{\phi^0 \phi^0 }^2 &=& -\frac{1}{ V_{331} v_{\phi} \left(V_{331}^2+v_{\phi}^2\right)} \Big[ - V_{331}^3 v_{\phi}^3 (\lambda_{1}+\lambda_{2}+2 \lambda_{3}) +\frac{\nu }{ \sqrt{2} } ( V_{331}^2 - v_{\phi}^2 )^2 v_u   \Big]   \,,\\[1mm]
M_{h_u  h_1^\prime}^2 &=& v_u s_{\tilde \beta} c_{\tilde \beta} (\kappa_{1}-\kappa_{2}) \sqrt{V_{331}^2+v_{\phi}^2}   \,,\\[1mm]
M_{ h_u h_2^\prime }^2 &=& \frac{v_u \Big[  s_{\tilde \beta}^2  (\kappa_{2} V_{331}^2-\kappa_{1} v_{\phi}^2 )+c_{\tilde \beta}^2  (\kappa_{1} V_{331}^2-\kappa_{2} v_{\phi}^2 ) \Big] }{\sqrt{V_{331}^2+v_{\phi}^2}}   \,,\\[1mm]
M_{\phi^0 h_1^\prime }^2 &=& V_{331} v_{\phi} s_{\tilde \beta} c_{\tilde \beta} (\lambda_{2}-\lambda_{1})   \,,\\[1mm]
M_{\phi^0  h_2^\prime }^2 &=& \frac{1}{2} \Big[ V_{331} v_{\phi} c_{2\tilde \beta} (\lambda_{2}-\lambda_{1}) - \frac{(V_{331}^2 -v_{\phi}^2 ) \left(V_{331} v_{\phi} (\lambda_{1}+\lambda_{2}+2\lambda_{3})+2 \sqrt{2} \nu v_u\right)}{V_{331}^2+v_{\phi}^2} \Big]   \,,\\[1mm]
M_{h_1^\prime h_1^\prime}^2 &=& -\frac{1}{8 V_{331} v_{\phi}}\left(V_{331}^2+v_{\phi}^2\right) (V_{331} v_{\phi} c_{4\tilde \beta} (\lambda_{1}+\lambda_{2}-2 (\lambda_{3}+\lambda_{4}+\lambda_{5})) \non
&-& V_{331} v_{\phi} (\lambda_{1}+\lambda_{2}+2 (-\lambda_{3}+\lambda_{4}+\lambda_{5}))+4 \sqrt{2} \nu v_u)   \,,\\[1mm]
M_{h_1^\prime  h_2^\prime }^2 &=& \frac{1}{4} s_{2\tilde \beta} \Big[ c_{2\tilde \beta} \left(V_{331}^2+v_{\phi}^2\right) (\lambda_{1}+\lambda_{2}-2 (\lambda_{3}+\lambda_{4}+\lambda_{5})) +(\lambda_{1}-\lambda_{2}) (V_{331}^2 -v_{\phi}^2 )  \Big]   \,,\\[1mm]
M_{h_2^\prime h_2^\prime }^2 &=& \frac{1}{8 \left(V_{331}^2+v_{\phi}^2\right)} \Big[ (V_{331}^4 + v_{\phi}^4) (3\lambda_{1}+3\lambda_{2}+2 (\lambda_{3}+\lambda_{4}+\lambda_{5}))+c_{4\tilde \beta} \left(V_{331}^2+v_{\phi}^2\right)^2 (\lambda_{1}+\lambda_{2}\non
&-&2 (\lambda_{3}+\lambda_{4}+\lambda_{5})) + 4 c_{2\tilde \beta} (\lambda_{1}-\lambda_{2}) \left(V_{331}^4-v_{\phi}^4\right)\non
&-& 2 V_{331}^2 v_{\phi}^2 (\lambda_{1}+\lambda_{2}+6 \lambda_{3}-2 (\lambda_{4}+\lambda_{5})) - 16 \sqrt{2} \nu  V_{331} v_{\phi} v_u) \Big]   \,.
\eeqn
\eeqs

\subsection{The Yukawa couplings of fermions}

In terms of the fermion mass eigenstates of $(b^\prime \,, B^\prime )$, the CP-odd and CP-even Higgs Yukawa couplings are expressed as follows
\beqs
\beqn
-\Lc_Y^{\Qc\,, 0^-}&\supset&  i m_b  \overline{b_L^\prime }  b_R^\prime  \Big(  \zeta_1 \eta_1 +  \zeta_2 \eta_2  +  \zeta_3 \pi_1 + \zeta_4  \pi_2 \Big)  \non
&+& i m_B \overline{b_L^\prime }  B_R^\prime \Big(  \zeta_1^\prime \eta_1 +  \zeta_2^\prime \eta_2  +  \zeta_3^\prime  \pi_1 + \zeta_4^\prime \pi_2 \Big) \non
&+& i m_b \overline{B_L^\prime }  b_R^\prime \Big( -\zeta_3 \eta_1 -\zeta_4  \eta_2 +  \zeta_1 \pi_1 + \zeta_2  \pi_2 \Big) \non
&+& i m_B  \overline{B_L^\prime }  B_R^\prime  \Big( -\zeta_3^\prime \eta_1 - \zeta_4^\prime \eta_2  + \zeta_1^\prime \pi_1 + \zeta_2^\prime \pi_2 \Big) +  H.c.\,,\label{eq:bbYuk_CPodd}\\[1mm]
-\Lc_Y^{\Qc\,, 0^+}&\supset& m_b  \overline{b_L^\prime }  b_R^\prime  \Big(  \zeta_1 \phi_1 + \zeta_2 \phi_2  + \zeta_3 h_1 + \zeta_4  h_2 \Big) \non
&+& m_B \overline{b_L^\prime }  B_R^\prime \Big( \zeta_1^\prime \phi_1 +  \zeta_2^\prime  \phi_2  +  \zeta_3^\prime  h_1 + \zeta_4^\prime  h_2 \Big)\non
&+& m_b \overline{B_L^\prime }  b_R^\prime \Big(  -\zeta_3 \phi_1 - \zeta_4  \phi_2  +  \zeta_1 h_1 + \zeta_2  h_2 \Big) \non
&+& m_B  \overline{B_L^\prime }  B_R^\prime  \Big(  -\zeta_3^\prime \phi_1 - \zeta_4^\prime \phi_2  +  \zeta_1^\prime  h_1 + \zeta_2^\prime  h_2 \Big) +  H.c.  \,,\label{eq:bbYuk_CPeven}
\eeqn
\eeqs
where we parametrize the couplings as follows
\beqs
\beqn
&& \zeta_1= c_L^2 \frac{s_{\tilde \beta} }{ v_\phi } - s_L c_L \frac{c_{\tilde \beta} }{ V_{331} } \,,\quad \zeta_1^\prime =  s_L c_L \frac{s_{\tilde \beta} }{ v_\phi } +  c_L^2 \frac{c_{\tilde \beta} }{ V_{331} }  \,,\\
&& \zeta_2 = - c_L^2 \frac{ c_{\tilde \beta} }{ v_\phi } - s_L c_L \frac{s_{\tilde \beta} }{ V_{331} } \,,\quad   \zeta_2^\prime =  - s_L c_L \frac{ c_{\tilde \beta} }{ v_\phi } +  c_L^2 \frac{s_{\tilde \beta} }{ V_{331} } \,,\\
&& \zeta_3 =  -s_L c_L \frac{s_{\tilde \beta}  }{ v_\phi } + s_L^2 \frac{c_{\tilde \beta} }{ V_{331} } \,,\quad \zeta_3^\prime = -s_L^2 \frac{s_{\tilde \beta} }{ v_\phi } - s_L c_L  \frac{c_{\tilde \beta} }{ V_{331} } \,,\\
&& \zeta_4 =  s_L c_L \frac{c_{\tilde \beta} }{ v_\phi } + s_L^2 \frac{ s_{\tilde \beta}  }{ V_{331} } \,,\quad  \zeta_4^\prime = s_L^2 \frac{c_{\tilde \beta} }{ v_\phi } - s_L c_L \frac{ s_{\tilde \beta} }{ V_{331} } \,.
\eeqn
\eeqs
For the fermion mass eigenstates of $( \tau^\prime \,, E^\prime )$, their Yukawa couplings can be obtained by replacing $(m_b\,, m_B)\to ( m_\tau \,, - m_E)$ in the above Eqs.~\eqref{eq:bbYuk_CPodd} and \eqref{eq:bbYuk_CPeven}.
\bigskip

\bibliographystyle{utphys.bst}

\providecommand{\href}[2]{#2}\begingroup\raggedright\endgroup

\end{document}